\documentclass{aa}  

\usepackage{graphicx}
\usepackage[usenames,dvipsnames]{xcolor}
\usepackage{natbib}
\usepackage{multirow}
\usepackage{booktabs,caption}
\usepackage[flushleft]{threeparttable}
\usepackage{txfonts}
\usepackage[backref=page, hyperfootnotes=true, hidelinks, colorlinks, citecolor=blue, linkcolor=blue, linktocpage, bookmarks=true]{hyperref}
\usepackage{footnote}
\usepackage{float}
\usepackage[bottom, flushmargin]{footmisc}
\usepackage{dcolumn}
\newcolumntype{d}[1]{D{.}{.}{#1}}

\makeatletter
\newcommand*{\rom}[1]{\expandafter\@slowromancap\romannumeral #1@}
\makeatother

\def\teff{$T_{\rm eff}$}
\def\logg{$\log g$}
\def\kms{km\,s$^{-1}$}

\begin{document} 

   \title{Pushing Least-Squares Deconvolution to the next level: application to binary stars\thanks{The code is publicly available here: \url{https://github.com/AndrewStSp/LSDBinary}.}}

   \author{A. Tkachenko
          \inst{1}
          \and
          V. Tsymbal\inst{2}
          \and
          S. Zvyagintsev\inst{2}
          \and
          H. Lehmann\inst{3} 
          \and
          F. Petermann\inst{3}
          \and
          D.~E. Mkrtichian\inst{4}
          }

   \institute{Institute of Astronomy, KU Leuven, Celestijnenlaan 200D, B-3001 Leuven, Belgium\\
              \email{andrew.tkachenko@kuleuven.be}
         \and
Institute of Astronomy, Russian Academy of Sciences, 119017, Pyatnitskaya str., 48, Moscow, Russia
        \and
Th\"{u}ringer Landessternwarte Tautenburg, Sternwarte 5, 07778, Tautenburg, Germany
        \and
National Astronomical Research Institute of Thailand, 260 Moo 4, T. Donkaew, A. Maerim, Chiangmai, 50180 Thailand
             }

   \date{Received September 15, 1996; accepted March 16, 1997}

 
  \abstract
   {Eclipsing, spectroscopic double-lined (SB2) binaries remain to be the prime source of precise and accurate fundamental properties of stars. Furthermore, high-cadence spectroscopic observations of the eclipse phases allow us to resolve the Rossiter-McLaughlin effect whose modelling offers the means to probe spin-orbit misalignment in binaries.}
   {We aim to develop a method that provides precise and accurate measurements of radial velocities of both binary components, including the in-eclipse orbital phases where line profiles are subject to large distortions due to the Rossiter-McLaughlin effect. We also intend to separate spectral contributions of the primary and secondary components in the velocity space such that a time-series of the separated spectroscopic signals can be obtained throughout the binary orbit, preserving any line-profile variability (LPV) that might be present in either or both of those spectroscopic contributions.}
   {In this study, we provide a generalisation of the Least-Squares Deconvolution (LSD) method to SB2 systems. Our {\sc LSDBinary} algorithm is capable of working with both in-eclipse and out-of-eclipse spectra as input and delivers the LSD profiles, LSD-based model spectra, and precise RVs of both binary components as output. We offer an option to account for the Rossiter-McLaughlin effect in the calculation of the initial guess LSD profiles and components' flux ratio such that the effect can be modelled within the algorithm itself. In that case, the algorithm delivers both the LSD profiles and RVs that are no longer distorted by the Rossiter-McLaughlin effect. Otherwise, when geometry of the Rossiter-McLaughlin effect is ignored in the calculation of the initial guess, the {\sc LSDBinary} algorithm delivers a RV curve that contains contributions from both orbital motion of the star and spectral line distortions due to the Rossiter-McLaughlin effect.}
   {In this study, we provide an extensive test of the {\sc LSDBinary} software package on simulated spectra of artificial binaries resembling Algol-type systems and detached binaries with similar components. We study the effects of signal-to-noise-ratio of input spectra, resolving power of the instrument, uncertain atmospheric parameters of stars, and orbital properties of the binary system on the resulting LSD profiles and RVs measured from them. We find that atmospheric parameters have negligible effect on the shape of the computed LSD profiles while affecting mostly their global scaling. High-resolution ($R\gtrsim$60\,000) spectroscopic observations are required to investigate the Rossiter-McLaughlin effect in detail though a medium resolving power of $R\approx$25\,000-30\,000 might suffice when the amplitude of the effect is large. Our results are barely sensitive to signal-to-noise ratio of the input spectra provided they contain sufficient number of spectral lines, such as in A-type stars and later. Finally, the orbital inclination angle and components' radii ratio are found to have the largest effect on the shapes of the LSD profiles and RV curves extracted from them.}
   {The {\sc LSDBinary} algorithm is specifically developed to perform detailed spectroscopic studies of eclipsing SB2 systems whose orbital configuration and components' atmospheric parameters are estimated by other means. The algorithm is well suited to study the Rossiter-McLaughlin effect as well as to compute the separated LSD profiles of both binary components from the observed composite in-eclipse spectra of SB2 systems.}


   \maketitle
%

\section{Introduction}

Stellar astrophysics has received an impressive boost with the launch and successful in-orbit operations of such space missions as the Microvariability and Oscillations of Stars \citep[MOST, ][]{Walker2003}, Convection, Rotation, and planetary Transits \citep[CoRoT,][]{Auvergne2009}, Kepler/K2 \citep[][]{Borucki2010,Howell2014}, and Transiting Exoplanet Survey Satellite \citep[TESS][]{Ricker2015}. While the first three missions have been retired in the meantime, TESS remains currently operational and holds potential to continue its science operations for as long as until 2030-2035. Furthermore, the PLAnetary Transits and Oscillations of stars \citep[PLATO, ][]{Rauer2014} European Space Agency mission is being planned for lunch in the fall 2026 and will continue the tradition of collecting large amounts of nearly uninterrupted, long time-base, ultra-high precision and duty cycle space-based photometry of stars and stellar systems.
  
Space-based photometric observations of stars and their ensembles are often complemented with ground-based spectroscopic observations enabling stellar astrophysics studies of unprecedented level of precision. For example, it has recently been demonstrated with asteroseismology \citep{Aerts2010} that models of stellar interior structure and evolution have largely incomplete treatment of angular momentum transport for intermediate- to high-mass stars \citep[e.g.][]{Mosser2012,Beck2012,Cantiello2014,Kawaler2015,Aerts2017,Aerts2019,Aerts2021}. These findings have in turn driven new theoretical developments that concern physics of internal gravity waves \citep[e.g.][]{Rogers2013,Rogers2015,Fuller2015a} and internal magnetic fields \citep[e.g.][]{Fuller2015b,Fuller2019}.

Parallel to detailed asteroseismic studies of interior structure of stars, the field of binary stars and multiple systems has also received an unprecedented boost thanks to the above-mentioned space missions. Nowadays, binary stars of various orbital and stellar type configurations are being found and observed in numbers, many of those also contain intrinsically variable stellar components \citep[e.g.,][]{Slawson2011,Matijevic2012,Conroy2014,Kirk2016,Sekaran2020,Ijspeert2021}. Co-existence of intrinsic variability of stars and binarity enables powerful synergies, where in particular asteroseismic studies benefit from knowledge of highly precise fundamental properties of stars that come from binary dynamics \citep[e.g.,][]{Schmid2015,Schmid2016,Johnston2019,Sekaran2021} while new aspects of tidal evolution theory can be developed and observationally verified thanks to interactions between stellar oscillations, magnetism, and binarity. This way, theory of tidal excitation of stellar pulsations could recently be tested and improved thanks to discovery and detailed observational studies of tidally-induced pulsations in close binary systems \citep[e.g.,][]{Welsh2011,Fuller2012,Fuller2013,Fuller2017}. At the same time, the effects of tides on self-excited stellar pulsations are also being actively studied and their interactions are being quantified \citep[e.g.,][]{Bowman2019,Handler2020,Fuller2020,Fuller2021}. Furthermore, it has been recently shown that observational incidence of global magnetic fields of presumably fossil origin among intermediate- to high-mass isolated stars amounts to some 5-10\% \citep[e.g.,][]{Alecian2019,Mathys2017}. Yet, there is a dearth of magnetic intermediate- to high-mass stars in close, short-period binaries, which is now claimed to be due to dissipation of the fossil magnetic fields by the turbulent magnetic diffusion induced by the saturated tidal flows \citep{Vidal2019}.

Besides serving as natural laboratories to study physics of interactions, binary stars are essential objects in galactic, stellar evolution, and distance scale context. First and foremost, the faction of stars found in binary and higher order multiple systems ranges from some 40\% for solar-like stars \citep{Cohen2020} and up to some 55-60\% for intermediate- to high-mass stars \citep[e.g.,][]{Almeida2020,Bodensteiner2021,Banyard2022}, hence any galactic-scale study that relies on the stellar component has to take the phenomenon of binarity into consideration. Furthermore, spectroscopic double-lined (SB2) eclipsing binaries are a prime source of precise and accurate fundamental properties of stars. With the stellar masses and radii being measured with accuracy better than 3\% \citep[e.g.,][]{Torres2010,Tkachenko2014,Debosscher2013,Pavlovski2009,Pavlovski2014,Pavlovski2018,Southworth2013,Southworth2021}, such binary systems provide one of the most stringent observational tests of stellar structure and evolution models. This way, it has been demonstrated that low-mass stars in binaries tend to have radii that inflated by some 5-15\% compared to the respective non-magnetic model predictions \citep[e.g.,][]{Ribas2006}. Inclusion of the magnetic fields in these models helps to resolve the issue \citep[e.g.,][]{MacDonald2017} demonstrating the efficiency of binaries in probing and improving stellar models in the regime of low-mass (M-type) stars. Similarly, binarity is being exploited to test interior physics in the models of intermediate- to high-mass stars in a way that is independent of and complementary to the method of asteroseismology. This way, several studies offer probes of the level of core-boundary mixing in the form of the convective core overshooting \citep[][]{Guinan2000,Claret2016,Claret2017,Claret2018,Claret2019,Martinet2021}, while others go as far as to report the pertinent need for higher convective core masses in models of intermediate- to high-mass stars \citep[e.g.,][]{Tkachenko2020,Johnston2021}. In addition, interferometric eclipsing SB2 systems provide the means of measuring their accurate dynamical parallaxes, hence offering an important reference for astrometric missions such as Hipparcos and Gaia \citep[e.g.,][]{Gallenne2019,Pavlovski2022}. Last but not least, eclipsing (binary) and transiting (exoplanet) systems offer a unique opportunity to study the effects of spin-orbit (mis)alignment through observations and interpretation of the Rossiter-McLaughlin (RM) effect \citep[e.g.,][]{Rossiter1924,McLaughlin1924,Albrecht2012,Triaud2018,Wang2018,Addison2018,Kamiaka2019}.

With the amount, and more importantly unprecedented quality of (space-based) photometric and (ground-based) spectroscopic observations available nowadays, considerable time and effort are being spent to improve methods and techniques for the analysis of stars in binary and higher order multiple systems. For example, following the development of binary light curve analysis algorithms such as {\sc Wilson-Devinney} \citep{Wilson1971}, {\sc Phoebe} \citep{Prsa2005}, and {\sc jktebop} \citep{Southworth2004} in the pre-space missions era, new methods such as those implemented in the {\sc ellc} \citep{Maxted2016} and {\sc Phoebe 2.0} \citep{Prsa2013,Degroote2013,Prsa2018,Conroy2019,Conroy2020,Conroy2021} codes are being developed, that are both more efficient and often include a better treatment of physics in them. In spectroscopy, the original method of cross-correlation \citep[e.g.,][]{Bracewell1965} has been commonly used in the past but proved impractical in many cases due to its strong dependence on the underlying template (synthetic) spectrum and inability to pick line profile variations intrinsic to the star (unless exceptionally strong, e.g. in large amplitude radially pulsating stars). Yet, two-dimensional cross-correlation analysis \citep[e.g., as implemented in the {\sc TODCOR} software package,][]{Mazeh1994} is often used for the RV determination in SB2 systems, despite the above-mentioned dependence on underlying model and number of degeneracies this method is subject to. A considerable improvement has been achieved with the introduction of the broadening function method \citep{Rucinski1992}, which is different from the original method of cross-correlation in that it uses a template that is not subject to spectral line broadening mechanisms such as stellar rotation and/or macroturbulence. This makes the method less model-dependent in the sense that small deviations of the template parameters from the true properties of the star are less critical for the overall shape of the obtained broadening function. In addition, the use of a template that incorporates only the intrinsic (e.g., thermal) broadening of spectral lines allows us to resolve fine structures in the ensemble of lines in the observed stellar spectrum and enables detection and further analysis of line profile variations caused by intrinsic variability of the star. 

The need to achieve extremely high signal-to-noise ratio (S/N) to be able to detected low-amplitude magnetic signatures in polarisation spectra has driven development of the Least-Squares Deconvolution \citep[LSD;][]{Donati1997} multi-line averaging technique. The LSD profiles obtained with this methods are also ideal for inference of radial velocities of stars (including those in binary systems) and to study temporal spectral changes due to intrinsic variability of stars. In this paper, we present a generalisation of the original LSD method to binary stars. In particular, we focus on the determination of precise, orbital phase-resolved radial velocities (RV) of the individual binary components with a careful treatment of the Rossiter-McLaughlin effect. In Section~\ref{LSD_method}, we provide a short introduction to the original LSD method as well as the literature overview with respect to the recent improvements and generalisations of the method. Our own generalisation of the method to spectroscopic double-lined binaries is described in detail in Section~\ref{Sect:LSDBinary}. The method is further tested and validated on artificial spectra of SB2 stars in Section~\ref{LSD_test_synthetic}, where we also touch upon its applicability range and limitations. We close the paper with conclusions and an outline of future prospects in Section~\ref{Conclusions}.

\section{The least-squares deconvolution (LSD) method}\label{LSD_method}

The LSD method was originally introduced by \citet{Donati1997} with the purpose of detection and measurement of weak surface magnetic fields from polarised spectra of stars. In its original formulation, the method relies on the following two fundamental assumptions: 1) all spectral lines in the considered spectral interval have similar profile shape, and 2) all lines in the spectrum add up strictly linear. Assuming both conditions are adequate for representing a stellar spectrum, the least-squares deconvolution problem can be formulated as a convolution of an a priori unknown {\it average profile} $Z(\upsilon)$ with a pre-computed {\it line mask} $M$
\begin{equation}
I = M*Z(\upsilon),
\end{equation}
where $\upsilon$ and $I$ stand for the velocity and model spectrum, respectively, while the line mask $M$ represents a set of delta functions (wavelength vs. predicted central line depths). The concept of representing observed spectrum as a convolution of a priori unknown average profile with a list of delta functions allows to boost S/N significantly. The exact amount of the S/N boost depends on spectral type of the star and wavelength range employed for the calculation of LSD profiles, and is typically anywhere between a factor of 5 to 50. Such a large increase in S/N of the data in turn enables the detection of stellar surface magnetic fields as week as a few tens of Gauss from the originally moderate S/N polarisation spectra \citep[e.g.,][]{Blazere2016a,Blazere2016b,Kochukhov2019}.

In practice, neither of the above two fundamental assumptions is fulfilled. It is a well known fact that strong (close to resonance) lines differ substantially in shape from their weak counterparts, and that lines with pronounced damping wings (e.g., Balmer lines in the spectra of main-sequence intermediate-mass stars and/or magnesium lines in solar-like stars) exhibit shapes that are very distinct from any other type of spectral lines in the spectrum. Furthermore, it is fair to assume that individual spectral lines add up linearly when blending occurs due to a superposition of two stellar spectra (i.e., the case of a double-lined spectroscopic binary) and/or due to a large rotational broadening. However, the above assumption is no longer valid when absorption coefficients of two or more neighbouring lines overlap in wavelength, which results in a non-linear addition of the lines forming a blend. Both these limitations give rise to non-negligible uncertainties in the predicted intensities of spectral lines in the {\it model spectrum I} obtained by a convolution of the {\it LSD profile Z($\upsilon$)} with the input {\it line mask M}. This implies that LSD profiles computed from a time-series of observed spectra can be studied with respect to time-dependent variability they exhibit, however each of these LSD profiles separately can hardly be interpreted as an isolated spectral line with properties averaged across the ensemble of spectral lines the LSD profile is computed from \citep{Kochukhov2010}. 

Several studies have suggested improvements to the original LSD technique with the ultimate goal to overcome major shortcomings of the method associated with its fundamental assumptions. \citet{Sennhauser2009} and \citet{Sennhauser2010} propose to use the Minnaert's analytical expression \citep{Minnaert1935} to account for non-linearity in spectral line blending when it comes to representation of strong, optically thick lines. The authors demonstrate that their method of nonlinear deconvolution with deblending ({\sc NDD}) offers a better representation of blended spectra than the original {\sc LSD} method that assumes linearity. \citet{Kochukhov2010} propose an improved method of least--squares deconvolution ({\sc iLSD}) which includes option of a {\it multiprofile} LSD that represents stellar spectrum as a superposition of an arbitrary number of scaled average profiles. The {\sc iLSD} method offers the opportunity of grouping spectral lines according to their predicted depths and representing each of the above groups with its own average profile. Given that weak and strong lines tend to show significant difference in their profile shapes, the approach proposed by \citet{Kochukhov2010} allows for partial compensation of one of the fundamental assumptions in the original formulation of the method, namely a shared profile shape between all spectral lines engaged in the calculation of the LSD profile. Furthermore, the authors demonstrate that the LSD profile cannot be interpreted as a real spectral line with average properties when it comes to Stokes~{\it I} (intensity) and Stokes~{\it Q} (linear polarisation) spectra. At the same time, the Stokes~{\it V} average profiles do resemble closely the behaviour of an isolated line with average properties, provided global magnetic field is weaker than some 1~kG. 

\citet{Tkachenko2013} elaborate further on the idea by \citet{Kochukhov2010} and implement in their approach a possibility to include multiple line masks $\sum_{i=1}^{N} M_i$ along with a multiprofile LSD. In particular, such approach allows one to compute composite average profiles from spectra of double-lined binary stars, taking into account difference in atmospheric parameters and chemical composition of the two binary components. Furthermore, the authors implement a line-strength correction algorithm that aims to improve upon the LSD representation of the stellar spectrum. Corrections are applied to the line strengths locally by minimising the differences between the observed and LSD-based model spectra. These corrections are pure mathematical and cannot be interpreted in terms of physics of stellar atmospheres, yet they allow for a significantly improved representation of the LSD-based model spectrum. The latter will have a much higher S/N than the original observed spectrum, hence the procedure by \citet{Tkachenko2013} offers an efficient way of denoising stellar spectra. 

Lastly, \citet{AsensioRamos2015} implement the {\sc LSD} algorithm under the Bayesian framework which enables fast calculation of the LSD average profile with a Gaussian prior. \citet{Strachan2017} introduce a method of differential least-squares deconvolution ({\sc dLSD}) that employs a high S/N combined spectrum of the star as a template and searches for a convolution kernel that needs to match the template to an observation of the star. The method is developed specifically to study planet-star obliquity through the interpretation of the observed Rossiter-McLaughlin effect.

\section{{\sc LSDBinary}: generalisation of the LSD method to spectroscopic double-lined binaries}
\label{Sect:LSDBinary}

The LSD method is by design sensitive to signals that are common to the majority of spectral lines in stellar spectrum. Therefore, even under the assumption of a single line mask $M$, LSD remains a powerful method for detection of SB2 binary systems. For example, the LSD profile computed from the composite spectrum of a SB2 system and employing a line mask corresponding to the primary component, will contain signatures of both binary components with a caveat that contribution of the secondary component to the LSD profile will be somewhat diluted. Measuring RVs and/or studying line profile variations (LPV) of individual binary components in these composite LSD profile still proves nearly as difficult as from the original observed composite spectra because, despite the significantly enhanced S/N in the LSD profile, line blending remains the dominant source of uncertainty. In practice, this means that the original LSD method as well as its numerous generalisations outlined in Section~\ref{LSD_method}, despite being effective for the {\it detection} of SB2 systems, are however not suitable for the calculation of LSD profiles of the individual binary components and inference of their precise RVs thereof.

In this paper, we present a generalisation of the original LSD method to SB2 binary star systems. In particular, we focus on separation of spectral contributions of the individual binary components in velocity space to the level of precision that would enable inference of precise and accurate RVs of both stars and, if possible, resolving their LPVs. In doing so, we focus primarily on the difficult case of the in-eclipse orbital phases that are characterised by high degree of blending, while we pay little attention to the out-of-eclipse phases where spectral contributions of the two stars can typically be easily separated from each other. This specific focus on the in-eclipse spectra is made solely for the purpose of demonstrating the algorithm's performance in the most difficult circumstances, while the algorithm is designed such as to handle all orbital phases in the spectroscopic observations of eclipsing SB2 systems. As a baseline, we adopt the modified LSD method by \citet{Tkachenko2013} that employs a modified, fast version of the Levenberg-Marquardt algorithm \citep{Marquardt1963} implemented by \citet{Piskunov2002}. The algorithm has the capabilities as outlined in the previous section and serves as the ``central engine'' of the {\sc LSDBinary} software package presented here.

\subsection{Pseudo-code}
\label{Sect:Pseudo-code}
\begin{figure*}
   \centering
   \includegraphics[width=18cm]{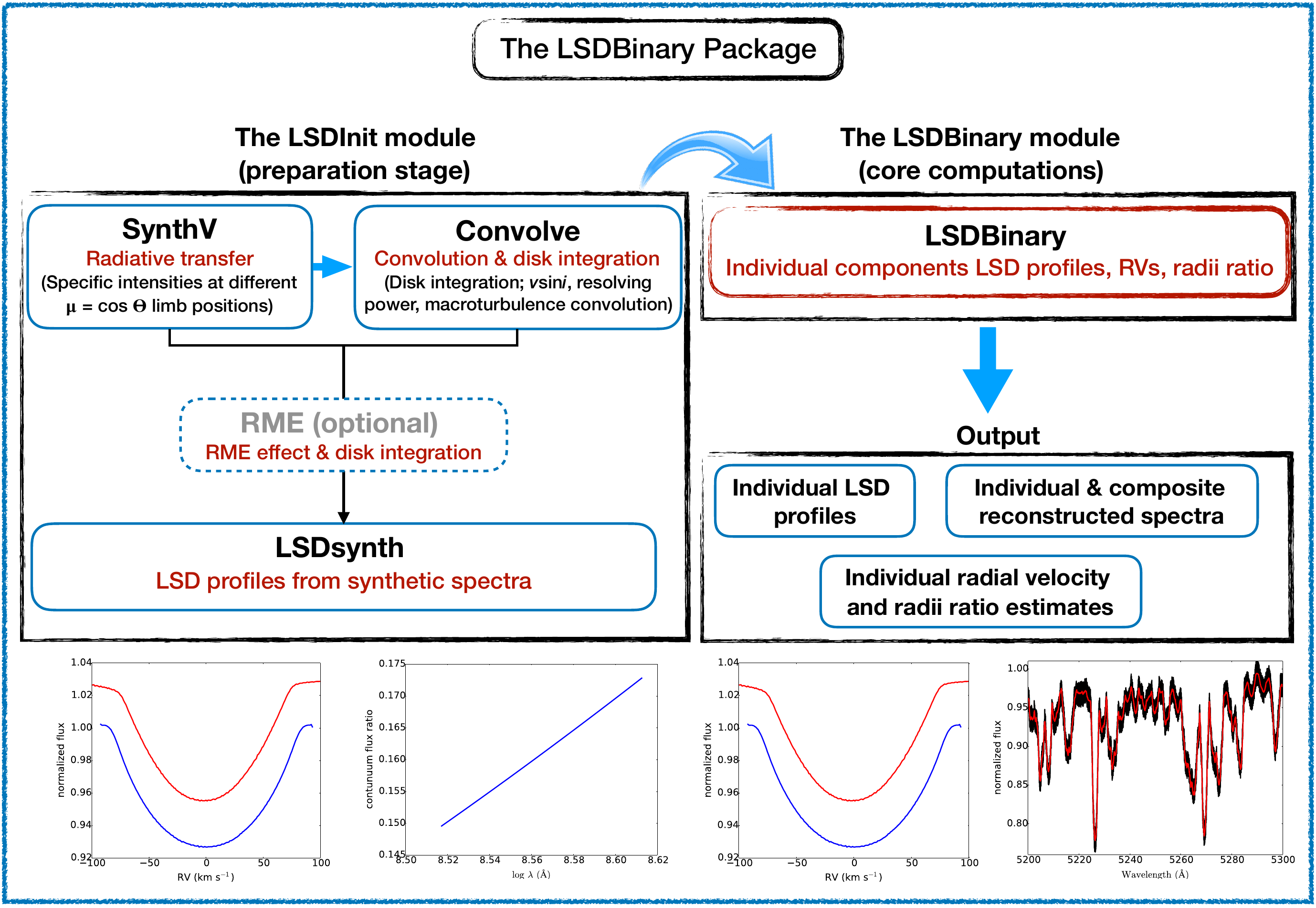}
      \caption{The {\sc LSDBinary} algorithm flow chart. Bottom left panels show output of the {\sc LSDinit} program in terms of the LSD profiles computed from synthetic spectra (left) and continuum flux ratio of the two binary components as a function of the wavelength (right). The output of the {\sc LSDBinary} program is illustrated in the two bottom right panels showing LSD profiles of the individual binary components (left) and a comparison between the composite observed (black) and LSD-based model (red) spectra of the binary. See Section~\ref{Sect:LSDBinary} for details.}
         \label{Fig:LSDBinary_FlowChart}
   \end{figure*}

Figure~\ref{Fig:LSDBinary_FlowChart} provides a flow chart of the {\sc LSDBinary} software package with the three main components - {\it input, central engine, {\rm and} output} - being indicated. Here, we provide a more detailed description of the entire process, in the form of a pseudo-code:
\begin{enumerate}
    \item The {\sc LSDinit} algorithm comprises three (optionally four) major steps (see below), whose purpose is to take care of all necessary preparations for the core calculations with the {\sc LSDBinary} algorithm. In this module, we set up initial guesses for the individual LSD profiles, wavelength-dependent flux ratio of the two stars, and pre-compute local corrections to the LSD model spectrum to ensure its closest match to the observations. The three (optionally four) above-mentioned steps are:
    \begin{itemize}
        \item The {\sc SynthV} radiative transfer code \citep{Tsymbal1996} is employed to compute synthetic spectra in an arbitrary wavelength range, with options for variable microturbulent velocity and chemical composition of the star, including possibility for their vertical stratification. Spectra are synthesised for different positions on the stellar disk to account for centre-to-limb intensity variations, also known as the limb darkening effect. The {\sc SynthV} code is publicly available and its latest version is distributed as part of the {\sc LSDBinary} package. However, {\sc SynthV} can also be replaced with any other one's favourite radiative transfer code provided its output is tuned to the input requirements of the {\sc LSDBinary} code.
        \item The {\sc Convolve} code \citep{Tsymbal1996} performs integration of specific intensities over the visible stellar disk and the convolution of the obtained spectrum with the Gaussian profile with the Full Width at Half Maximum (FWHM) corresponding to resolving power $R$ of the instrument, the projected rotational velocity $v\,\sin\,i$ of the star, and optionally the macroturbulent velocity. The macroturbulent broadening is implemented in the radial-tangential formalism following \citet{Gray1992}.
        \item The Rossiter-McLaughlin Effect ({\sc rme}) algorithm can be used as an alternative to calculation of the disk-integrated synthetic spectra with the {\sc SynthV} and {\sc Convolve} suit of codes and is therefore an optional step. The {\sc rme} algorithm, as the name suggests, allows us to compute time-series of the in-eclipse synthetic spectra subject to distortion caused by the Rossiter-McLaughlin effect \citep{Rossiter1924,McLaughlin1924} and the associated variable with orbital phase binary components' flux ratio. RME is a geometric effect that occurs when (stellar or planetary) companion transits across the disk of rotating primary component and blocks part of its visible surface. In doing so, the companion also blocks part of either blue (ingress) or red (egress)-shifted (due to stellar rotation) light of the primary component causing asymmetries in its observed line profiles. Our {\sc rme} algorithm solves the problem analytically by computing the surface area of the eclipsed star that is being blocked by the companion and performs the disk integration over its remaining visible part of the disk. The {\sc rme} algorithm relies on orbital configuration of the system as input and delivers distorted line profiles and orbital phase-dependent components' flux ratio to be used for the calculation of initial guess LSD profiles (see below) for the in-eclipse phases. Therefore, the option of using the {\sc rme} algorithm is best suitable for binary systems whose orbital configuration is know in advance, though the algorithm does not require the orbital parameters to be know with particularly high precision and/or accuracy.
        \item The {\sc LSDsynth} algorithm is designed to compute theoretical LSD profiles for both binary components from the corresponding disk-integrated synthetic spectra (with or without the RME taken into account) and line masks. To speed up calculations, the code employs as the initial guess for the LSD profile either a Gaussian profile with the FWHM corresponding to the resolving power $R$ of the instrument or a rotational kernel for the cases of slow and moderate to rapid rotation, respectively. An example of the synthetic LSD profiles computed with the {\sc LSDsynth} algorithm for both binary components is shown in Figure~\ref{Fig:LSDBinary_FlowChart} (bottom left panel). As discussed in Section~\ref{LSD_method}, due to the natal assumptions and limitations of the original LSD method, the LSD-based model spectrum $I$ computed by means of the convolution of the LSD profile $Z(v)$ with the line mask $M$ fails to closely reproduce the input stellar spectrum the LSD profile is computed from. The {\sc LSDsynth} algorithm developed by us overcomes that problem by computing and applying local corrections to line intensities in the LSD-based model spectrum such as to provide its closest match to the input synthetic stellar spectrum. These relative intensity corrections take the following form $(r_{\rm synthetic}-r_{\rm model})/r_{\rm model}$, where $r$ stands for the normalised flux, while ``synthetic'' and ``model'' refer to the input synthetic and LSD-based model stellar spectrum, respectively. Therefore, these local intensity corrections represent fractional differences between the input synthetic and LSD-based model spectra at each wavelength point, and ensure the closest match between the two when applied to the LSD-based model spectrum. Lastly, the {\sc LSDsynth} algorithm computes the ratio of binary component's fluxes as a function of wavelength and approximates it with the polynomial of second degree. An example of the functional form of this wavelength-dependent flux ratio is shown in Figure~\ref{Fig:LSDBinary_FlowChart} (bottom left panel). Summarising, taking synthetic spectra and line masks as input, the {\sc LSDsynth} algorithm delivers: (i) initial guess synthetic spectra-based LSD profiles, (ii) local fractional intensity corrections to the LSD-based model spectrum, and (iii) best fit polynomial coefficients that allow one to reproduce the wavelength-dependent functional form of the components' flux ratio.\\ 
    \end{itemize}
    \item The {\sc LSDBinary} algorithm represents the ``core engine'' of the method and allows us to compute time series of the LSD profiles of both binary components separated in velocity space. 
    \begin{itemize}
        \item {\it Input:} The algorithm relies on the output of the {\sc LSDinit} module in terms of the initial guess for: (i) the synthetic spectrum-based LSD profiles of both binary components, (ii) local fractional intensity corrections to the LSD-based model spectra of both stars, and (iii) functional form of the components' flux ratio. In addition to the above, the {\sc LSDBinary} algorithm also requires initial guess for the RVs of both binary components, that can be provided either in the form of a table, RV vs. orbital phase, or in the form of orbital parameters of the binary system. In the latter case, orbital phase resolved RVs are computed from the provided orbital elements within the {\sc LSDBinary} module and are employed as the initial guess. Lastly, {\sc LSDBinary} requires two line masks, one for each binary component, to be provided as well.
        \item {\it Core calculations:} {\sc LSDBinary} utilizes the LSD algorithm implemented in \citet{Tkachenko2013}, in particular its functionality to compute multi-component LSD profiles with the employment of multiple line masks. We modify the algorithm such as to account for relative contributions of the two binary components to the system's composite spectrum. These contributions are defined by the component's flux ratio whose wavelength-dependent functional form is pre-computed within the {\sc LSDinit} module and their radii ratio. The latter parameter can be either fixed (e.g., if it is known from the binary light curve solution) or used as a free parameter in {\sc LSDBinary}. The problem of solving for the individual LSD profiles is approached in iterative fashion with the employment of the modified version of the Levenberg-Marquardt algorithm \citep[see above and][]{Piskunov2002}. In each iteration, the algorithm minimises the difference between the input observed composite spectrum of the binary system and its LSD-based composite model spectrum counterpart. The latter is computed as a linear sum of the individual binary components' model spectra that are subject to their local fractional intensity corrections before being added up.
        \item {\it Output:} The {\sc LSDBinary} algorithm has Level-0 and Level-1 output. The former refers to the individual LSD profiles and the LSD-based model spectra. These model spectra are provided for: (i) both binary components separately and are the product of the convolution of the primary and secondary LSD profiles with their respective line masks, and taking the individual fractional intensity corrections into account, and (ii) the binary system as a whole, where the composite model spectrum is represented by the linear sum of the primary and secondary LSD-based model spectra with the components' flux and radii ratio being taken into account. On the other hand, the Level-1 output refers to the quantities inferred directly from the LSD profiles and the process of their calculation, i.e. from the Level-0 output. These quantities are the RVs of both binary components and, in case it is set as a free parameter, radii ratio of the two stars as a function of the orbital phase. Individual RVs are computed by matching the components' LSD profiles to the respective initial guess synthetic spectrum-based LSD profiles with the employment of the golden search algorithm \citep{Kiefer1953}. Therefore, RVs delivered by the {\sc LSDBinary} algorithm are on the scale relative to the initial guess LSD profiles. An example of the Level-0 output in terms of the individual LSD profiles and the LSD-based composite model spectrum overlaid on the input observed spectrum of the binary system is shown in Figure~\ref{Fig:LSDBinary_FlowChart} (bottom right panel)
    \end{itemize}   
\end{enumerate}

    \begin{figure*}
   \centering
   \includegraphics[width=5.8cm]{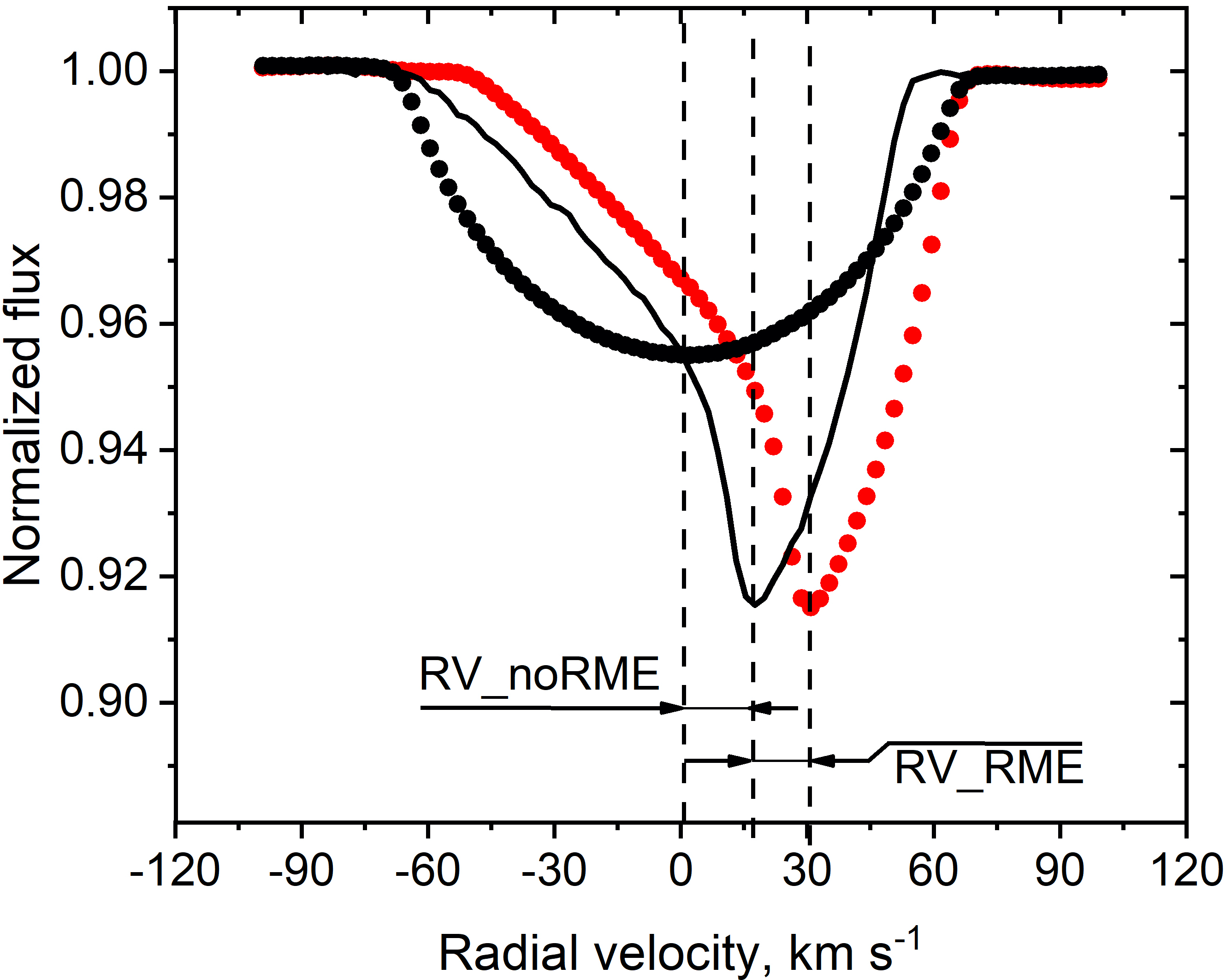}\hspace{2mm}
   \includegraphics[width=5.6cm]{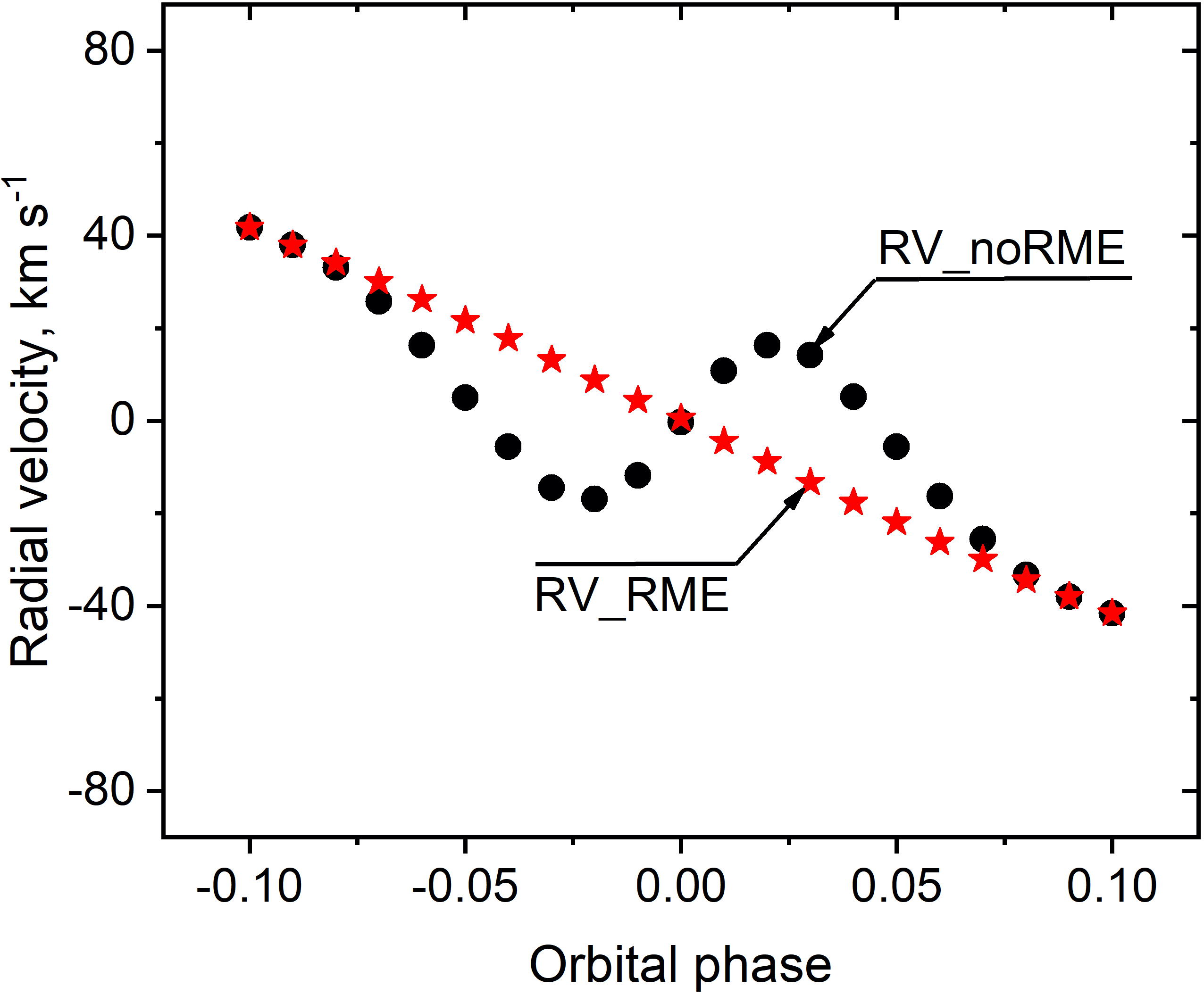}\hspace{2mm}
   \includegraphics[width=5.8cm]{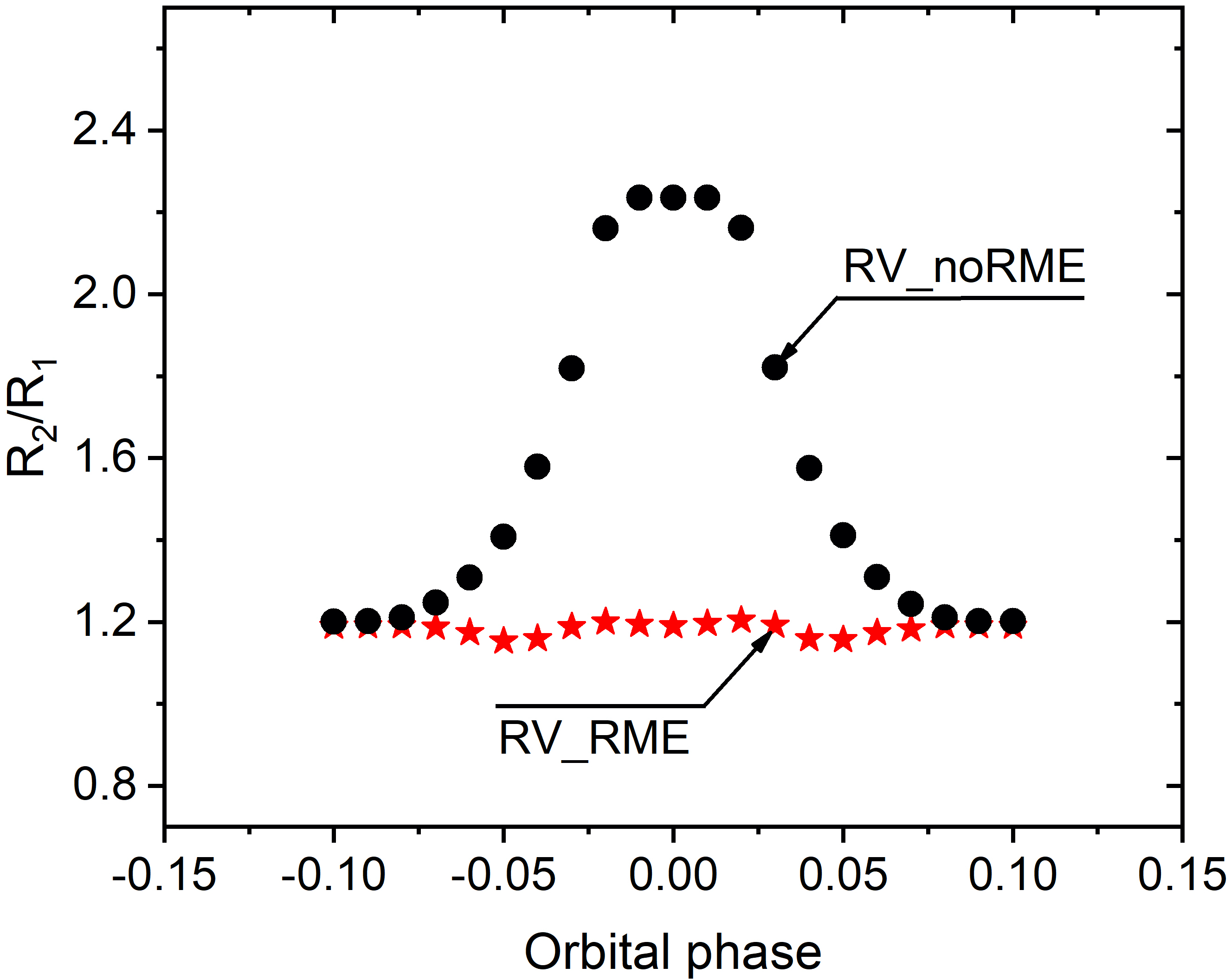}
      \caption{Graphical presentation of the {\sc LSDBinary} algorithm. {\bf Left:} initial guess synthetic LSD profiles computed with the {\sc LSDinit} module with (red stars) and without (black dots) the RM effect, along with the resulting ``observed'' LSD profile (black solid line) computed with the {\sc LSDBinary} module. {\bf Middle:} the in-eclipse phase resolved RV-curves with (red stars) and without (black dots) the RM effect. The RVs measured relative to the respective initial guess LSD profiles are indicated with the arrows. {\bf Right:} radii ratio as a function of orbital phase with (red stars) and without (black dots) the RM effect taken into account in the computation of the initial guess synthetic spectrum-based LSD profiles. Arrows indicate the same data points as in the middle panel. See text for details. 
              }
         \label{Fig:LSDBinary_demonstration}
   \end{figure*}

 \begin{figure*}
   \centering
   \includegraphics[width=17.5cm]{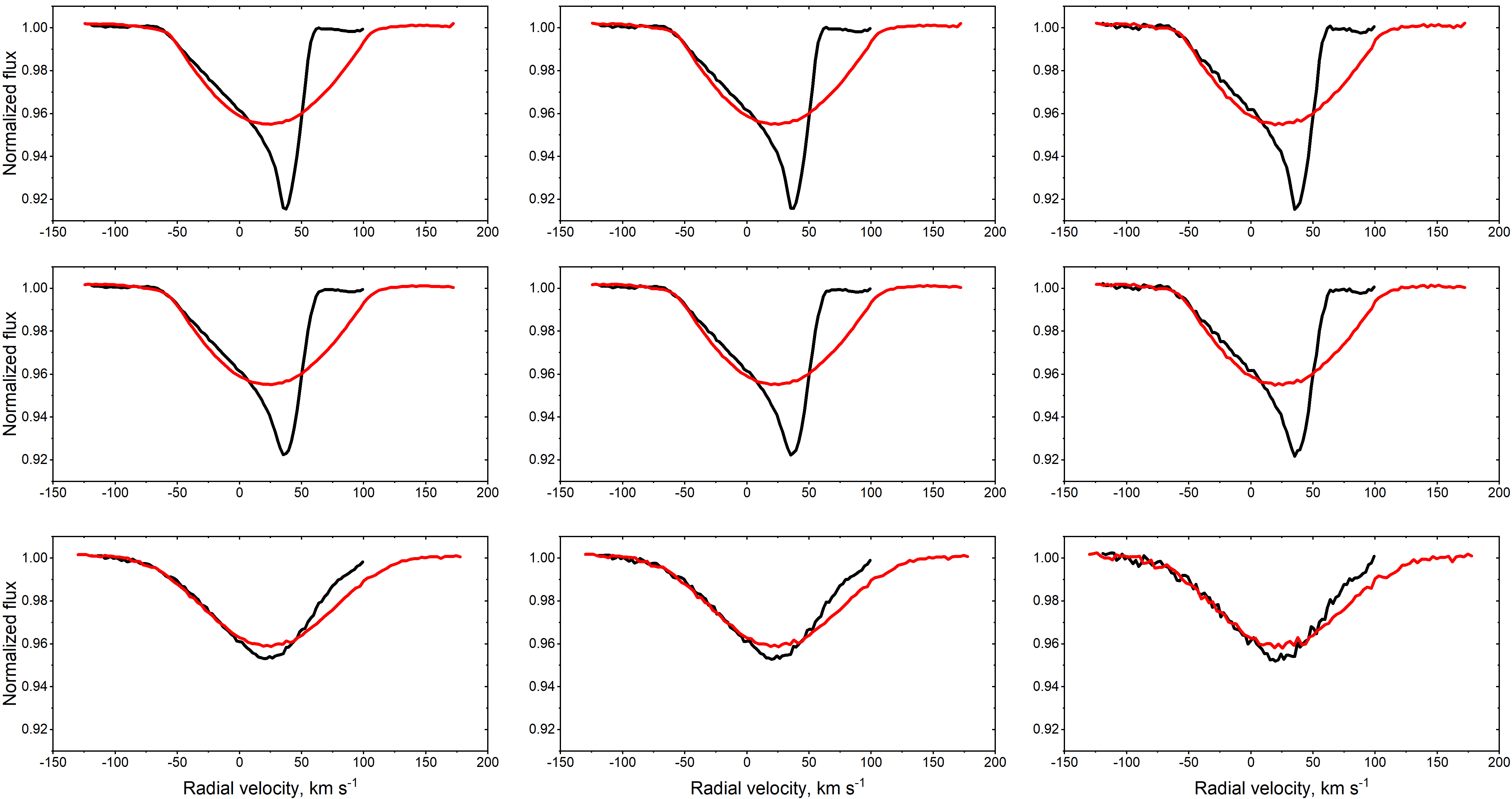}\vspace{10mm}
   \includegraphics[width=17.5cm]{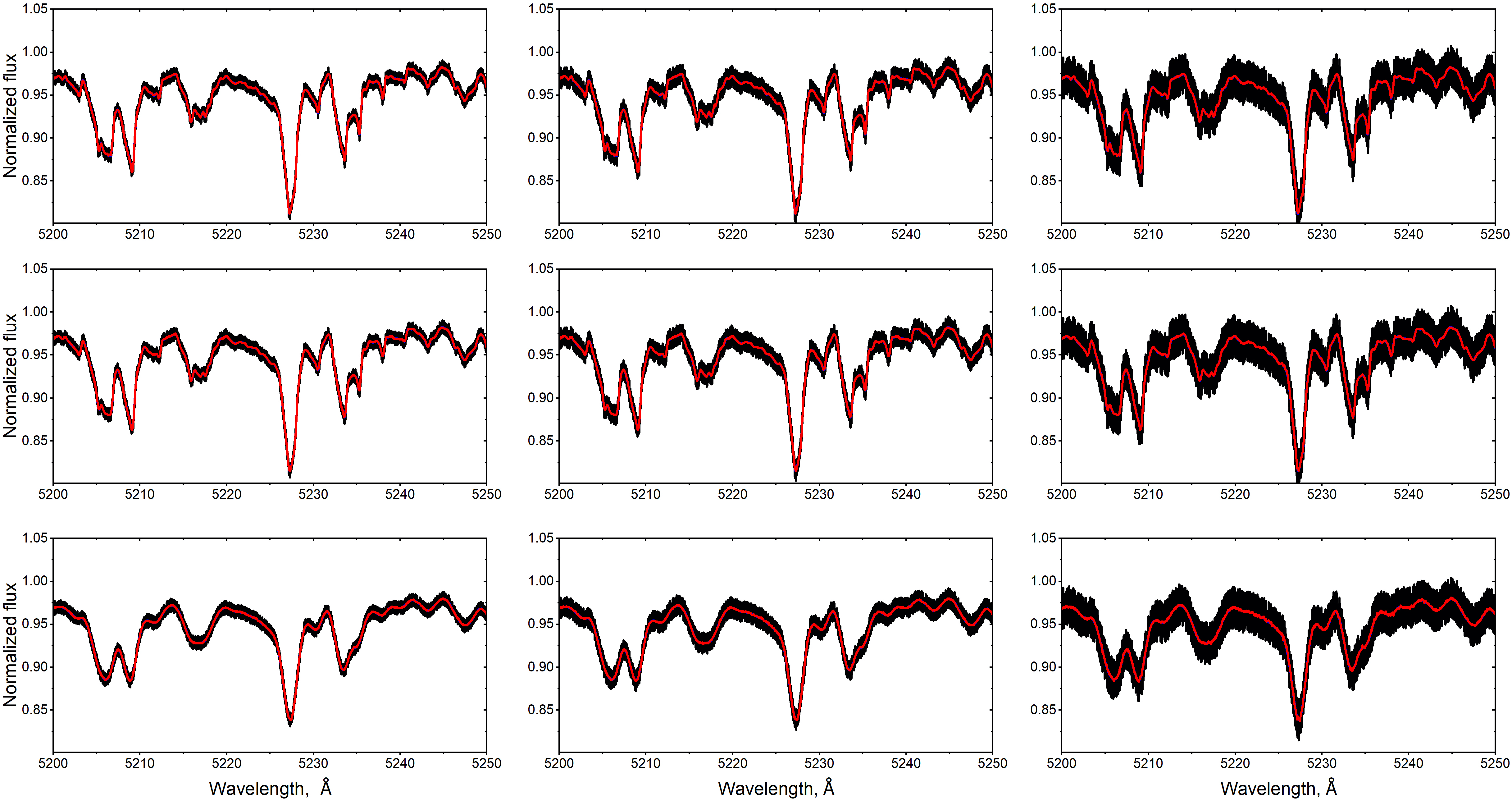}
      \caption{Results of the application of the {\sc LSDBinary} algorithm to artificial in-eclipse spectra of the RZ\,Cas binary system. {\bf Top block:} LSD profiles of the primary (red) and secondary (black) components. {\bf Bottom block:} LSD-based model spectra (red) overlaid on the input ``observed'' spectra (black) of the system. In each of the blocks, rows correspond to different values of the resolving power (from top to bottom, $R$=60\,000, 25\,000, and 5\,000) while columns reflect the change in S/N of the spectrum (from left to right, S/N=120, 80, and 40).}
         \label{Fig:LSDBinary_RZCas_artificial}
   \end{figure*}
   
   \begin{figure*}
   \centering
   \includegraphics[width=17.5cm]{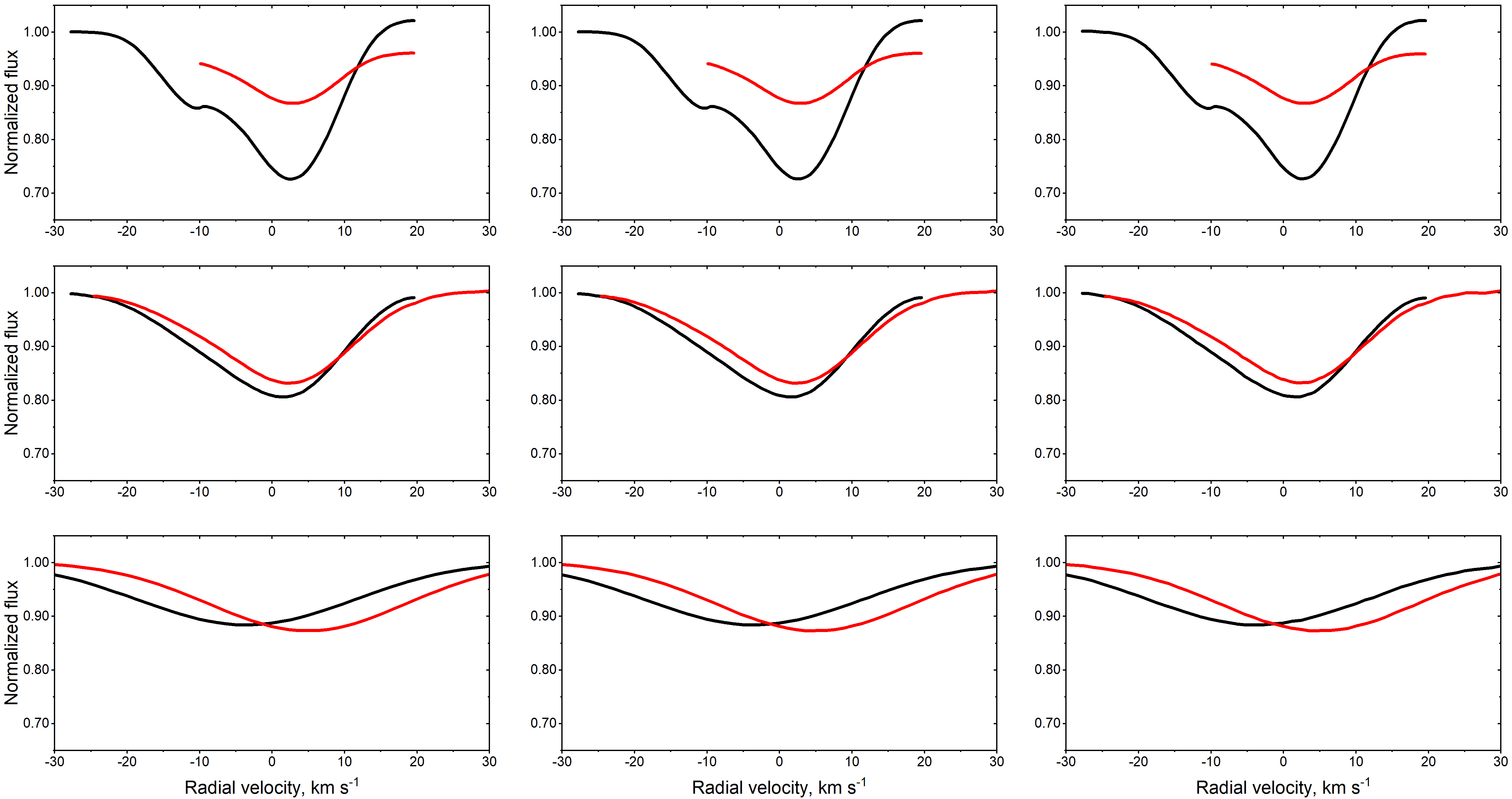}\vspace{5mm}
   \includegraphics[width=17.5cm]{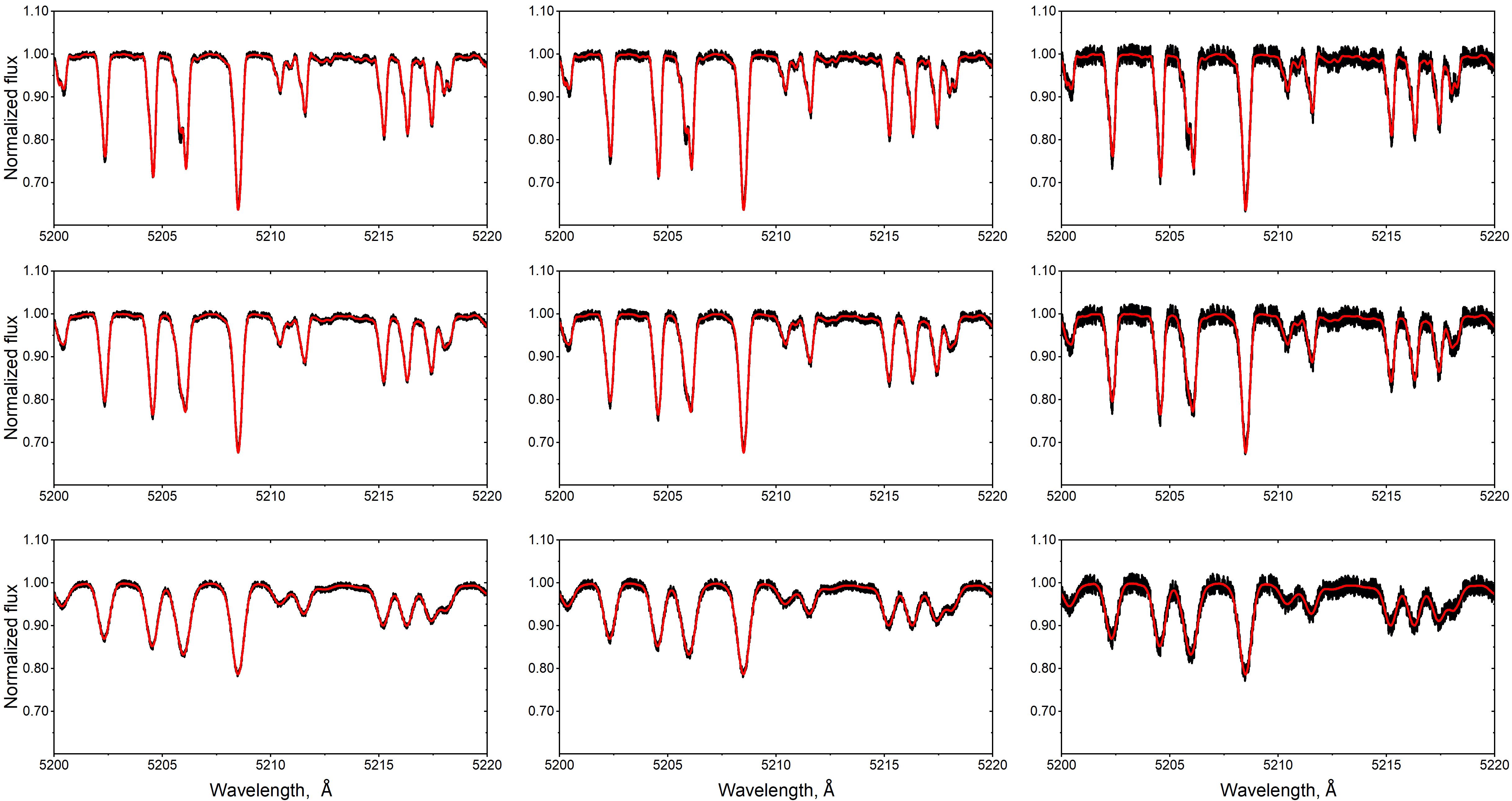}
      \caption{Same as Figure~\ref{Fig:LSDBinary_RZCas_artificial} but for the KIC\,11285625 system.
              }
         \label{Fig:LSDBinary_KIC11285625_artificial}
   \end{figure*}

\subsection{Step-by-step algorithm demonstration}
\label{Sect:Step-by-step_demonstration}
In order to provide a step-by-step demonstration of the {\sc LSDBinary} algorithm performance, we simulate the in-eclipse spectra of an Algol-like binary system with mid A-type primary and K-giant secondary components. Time-series of orbital phase resolved input spectra are simulated with our {\sc RME} module for a representative S/N of $\sim$100. In this ideal scenario, line masks are taken to fully correspond to the atmospheric parameters of the primary and secondary components. The initial guess, synthetic spectrum-based LSD profiles as well as the functional form of the wavelength-dependent flux ratio are computed with the {\sc LSDInit} module under the following assumptions: (i) LSD profiles of both stars are symmetric and their flux ratio is constant with orbital phase, and (ii) LSD profile of the primary component is distorted due to the RM effect and the components' flux ratio is orbital phase-dependent. In this particular example, as well as in the rest of this work, LSD profiles are computed from a 500~\AA\ wide wavelength interval centred at 5250~\AA. This wavelength interval is free of the Balmer lines and contains a large amount of metal lines in the spectra of A, F-type stars that we are concerned with in this work.

Left panel in Figure~\ref{Fig:LSDBinary_demonstration} shows a singleshot system's LSD profile (black solid line) computed with the {\sc LSDBinary} module at orbital phase $\phi=0.03$ ($\phi=0.0$ corresponds to the centre of the primary eclipse). The initial guess, synthetic spectrum-based symmetric (computed with the {\sc SynthV} and {\sc Convolve} suite of codes) and distorted by the RM effect (computed with the {\sc RME} module) LSD profiles are shown with the black dots and red stars, respectively. We stress that irrespective of the shape of the initial guess LSD profile, the output ``observed'' LSD profile is highly asymmetric and has the shape that is remarkably similar in both cases. In other words, the Level-0 data products of the {\sc LSDBinary} algorithm in the form of the system's LSD profile and the corresponding LSD-based model spectrum are barely sensitive to the assumed initial guess. Radial velocities computed from this singleshot system's LSD profile and measured relative to the respective initial guess synthetic spectrum-based LSD profiles (that are computed at laboratory wavelengths) are indicated with arrows in the middle panel in Figure~\ref{Fig:LSDBinary_demonstration} (designated as ``RV\_noRME'' and ``RV\_RME'' for the case of the symmetric and distorted initial guess profile, respectively). The middle panel also demonstrates the orbital phase-resolved RV curves for clarity of comparison between the two initial guess cases. We note that the two RV curves differ from each other substantially with one of them resembling pure orbital motion of the primary component (red symbols) whereas the other one also showing signatures of the line profile distortions due to the RM effect (black symbols). Because RV is an integral quantity of the line profile it is sensitive to both global shift of the profile due to the orbital motion of the star and the profile distortion, no matter what the true cause of that distortion is. In our example (see left panel in Figure~\ref{Fig:LSDBinary_demonstration}), RV of the ``observed'' LSD profile has positive difference of some 15~\kms\ with respect to the symmetric initial guess LSD profile. On the other hand, the same ``observed'' LSD profile has negative RV difference of some 10~\kms\ with respect to the asymmetric initial guess LSD profile that shares laboratory wavelengths with its symmetric version but now contains distortions due to the RM effect.

The effect of the initial guess LSD profile and the corresponding orbital phase-resolved functional form of the components' flux ratio on the inferred with {\sc LSDBinary} radii ratio parameter is illustrated in the right panel in Figure~\ref{Fig:LSDBinary_demonstration}. One can see that in the case of a distorted due to the RM effect initial guess LSD profile the radii ratio curve appears nearly flat (red stars). This result is indicative of {\sc LSDBinary} finding a constant with orbital phase value of the radii ratio that, in this particular case, also converges to the assumed in the simulations input value of $R_2/R_1=1.2$. This is because flux contribution of the primary component varies in the course of its eclipse and this variability is taken into account in the calculation of the initial guesses with the {\sc RME} module. The primary's spectrum is therefore modelled correctly in the {\sc LSDBinary} module. On the other hand, the assumption of a constant initial guess LSD profile and flux contribution of the primary component in the course of its eclipse leads to appreciable variations of the inferred with {\sc LSDBinary} radii ratio parameter (black dots). This result is indicative of incorrectly assumed orbital phase-independent flux contribution of the primary component. By assuming no flux variation for the primary component in the course of its obscuration by the secondary, we overestimate the primary's contribution to the composite spectrum of the binary system. The amount of this overestimation is variable and depends strongly on the surface area of the primary that is being blocked by the companion star. In practical terms, the assumption of a constant with orbital phase flux ratio leads to a significant overestimation of line depths for the primary and causes an appreciable mismatch between the observed and LSD-based model composite spectra of the binary system. In an attempt to compensate for this mismatch, the {\sc LSDBinary} module seeks to lower contribution of the primary component to the total flux of the system, the goal that is most efficiently achieved by increasing the radii ratio parameter $R_2/R_1$. The effect is expected to be more pronounced at phases where larger surface area of the primary is being blocked by the companion star, i.e. in the centre of the primary eclipse (just as we observe in the right panel in Figure~\ref{Fig:LSDBinary_demonstration}). 

In conclusion, though the Level-0 data products of the {\sc LSDBinary} algorithm (LSD profiles and LSD-based model spectra) are barely sensitive to the choice of the initial guess LSD profiles and constant versus variable flux ratio, its Level-1 data products (RVs and the radii ratio parameter) are influenced significantly by the choice made. In the next Section, we demonstrate how the sensitivity of the Level-1 data products to the choice of the initial guess can be exploited to improve upon the assumed orbital parameters of a binary system.

\section{{\sc LSDBinary}: method validation, applicability range, and limitations}\label{LSD_test_synthetic}
\label{Sect:LSDBinary_test_artificial}

\begin{table}
\begin{threeparttable}
\setlength{\tabcolsep}{4pt}
\caption{Summary of systemic and atmospheric parameters of the RZ\,Cas and KIC\,11285625 binary systems as derived in \citet{Tkachenko2013} and \citet{Debosscher2013}, respectively}
\label{Tab:RZCas_KIC11285625}      
\centering                          
\begin{tabular}{l c c c c}        
\toprule
\multirow{2}{*}{Parameter} & \multicolumn{2}{c}{RZ\,Cas} & \multicolumn{2}{c}{KIC\,11285625}\\
& Primary & Secondary & Primary & Secondary\\   
\midrule \hline
$T_{\rm eff}$ (K) & 8\,800 & 4\,800 & 7\,000 & 7\,200\\
$\log\,g$ (dex) & 4.3 & 3.5 & 4.0 & 4.2\\
$v\,\sin\,i$ (${\rm km}\,{\rm s}^{-1}$) & 65 & 80 & 14 & 8\\
${\rm [M/H]}$ (dex) & \multicolumn{2}{c}{0.0} & \multicolumn{2}{c}{0.0}\\
$K^{*}$ (${\rm km}\,{\rm s}^{-1}$) & 71.5 & 202 & 135 & 174\\
$P$ (days) & \multicolumn{2}{c}{1.195} & \multicolumn{2}{c}{10.79}\\
$i^{**}$ (degrees) & \multicolumn{2}{c}{82} & \multicolumn{2}{c}{85}\\
$e$ & \multicolumn{2}{c}{0.0} & \multicolumn{2}{c}{0.005}\\
$R_2/R_1$ & \multicolumn{2}{c}{1.19} & \multicolumn{2}{c}{0.69}\\
\bottomrule
\end{tabular}
\begin{tablenotes}
      \small
      \item $^{*}$ Radial velocity semi-amplitude
      \item $^{**}$ Orbital inclination angle
    \end{tablenotes}
\end{threeparttable}
\end{table}

\begin{figure*}[h!]
   \centering
    \includegraphics[width=8.5cm]{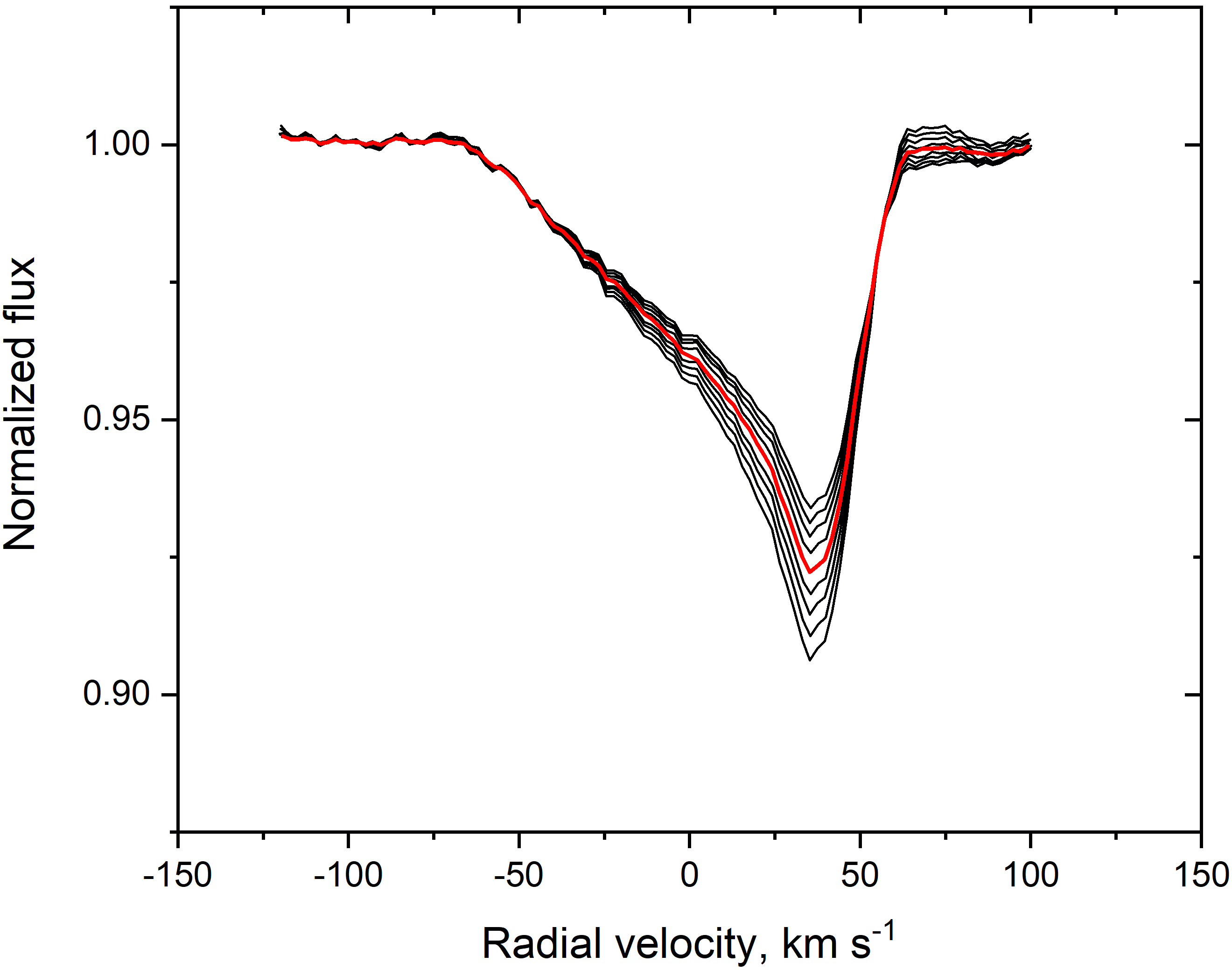}\hspace{5mm}
    \includegraphics[width=8.5cm]{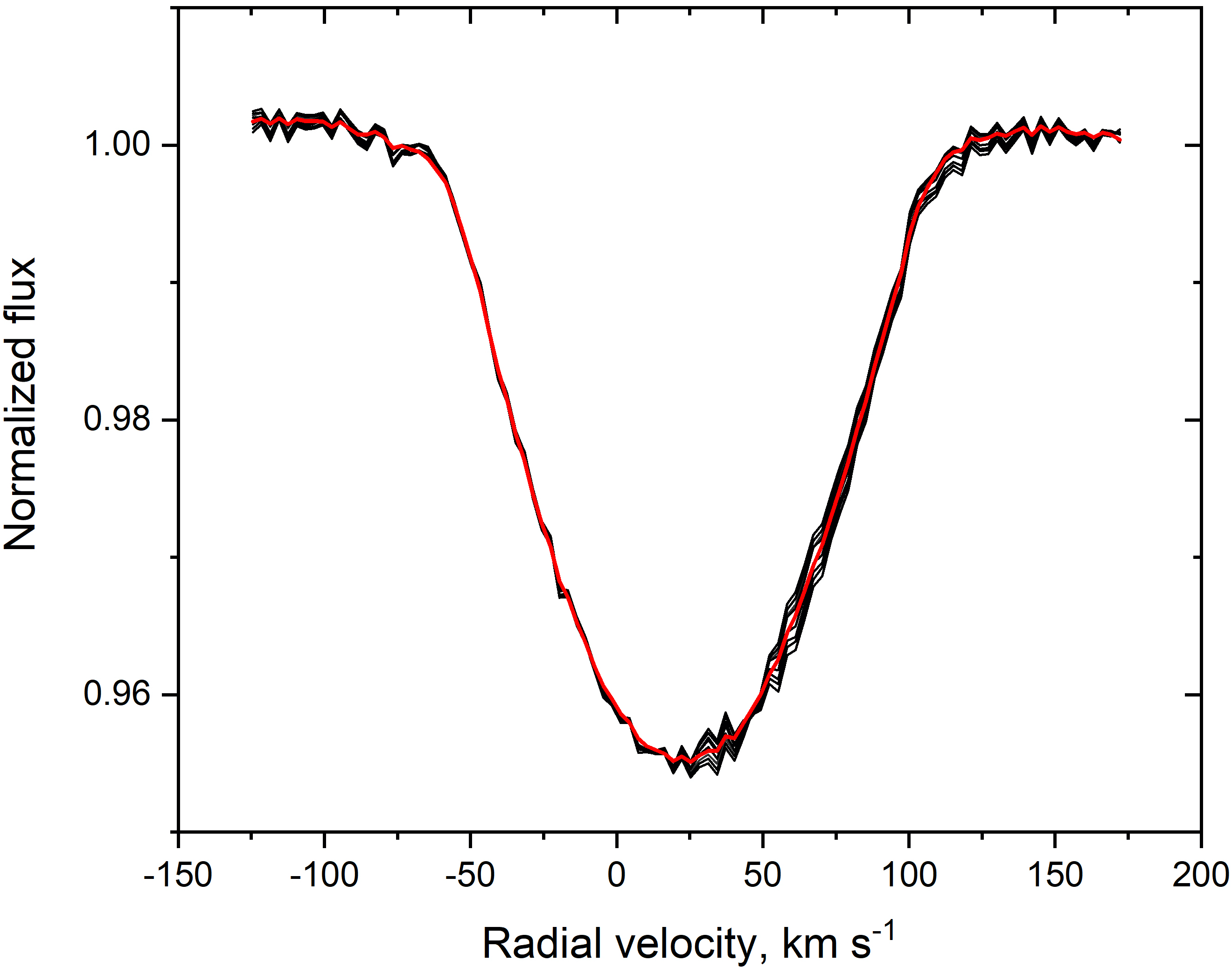}\vspace{5mm}
    \includegraphics[width=8.5cm]{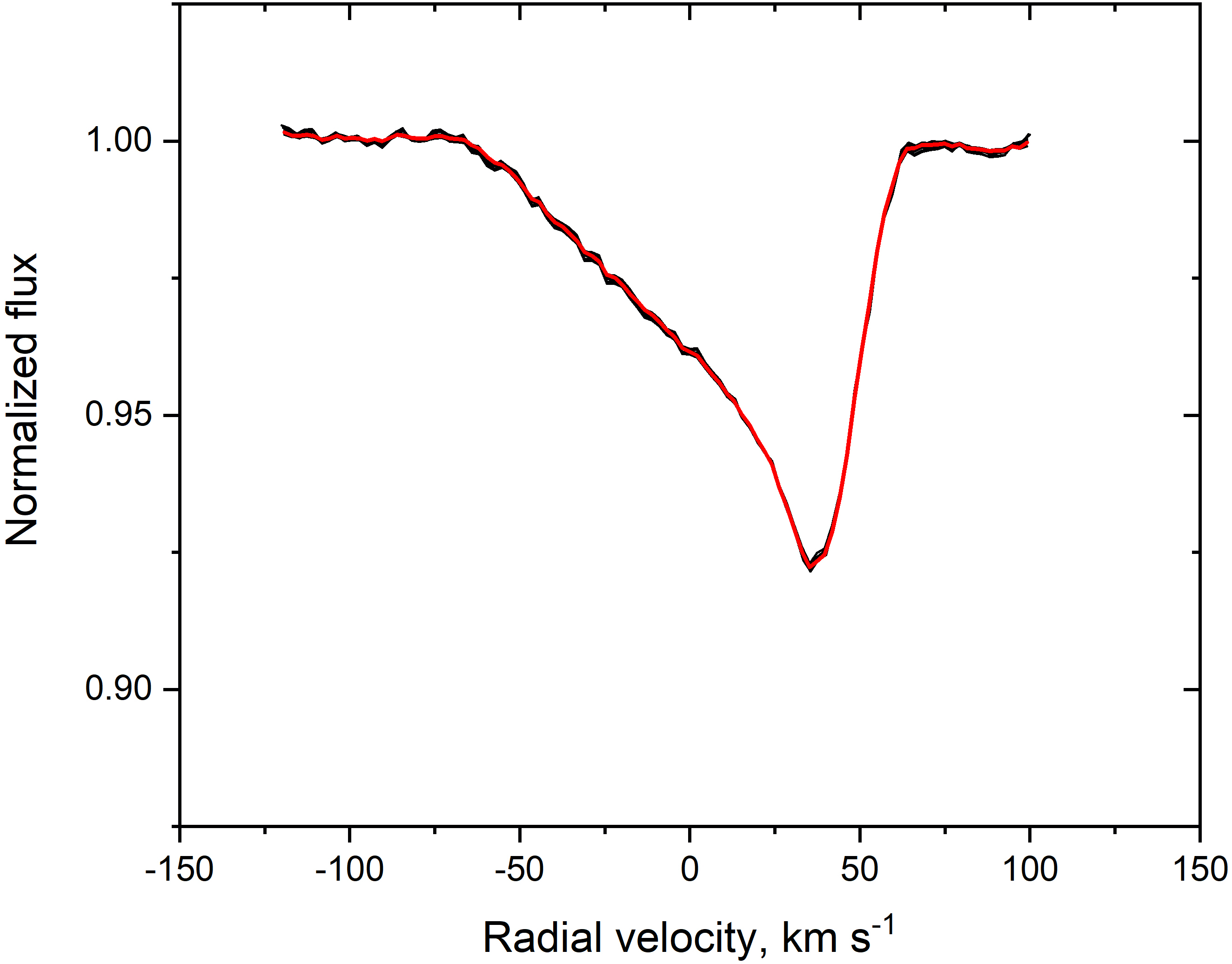}\hspace{5mm}
    \includegraphics[width=8.5cm]{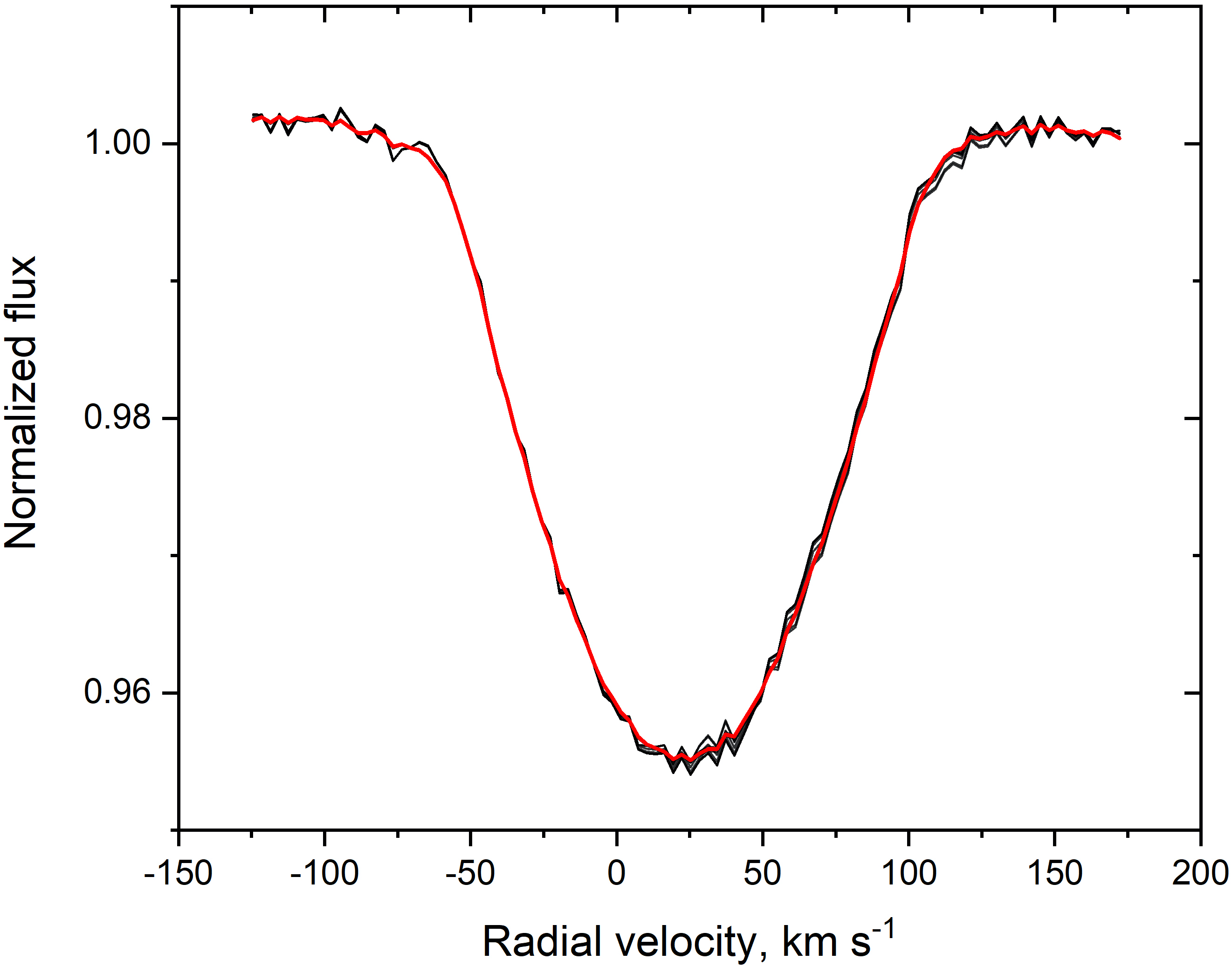}\vspace{5mm}
    \includegraphics[width=8.5cm]{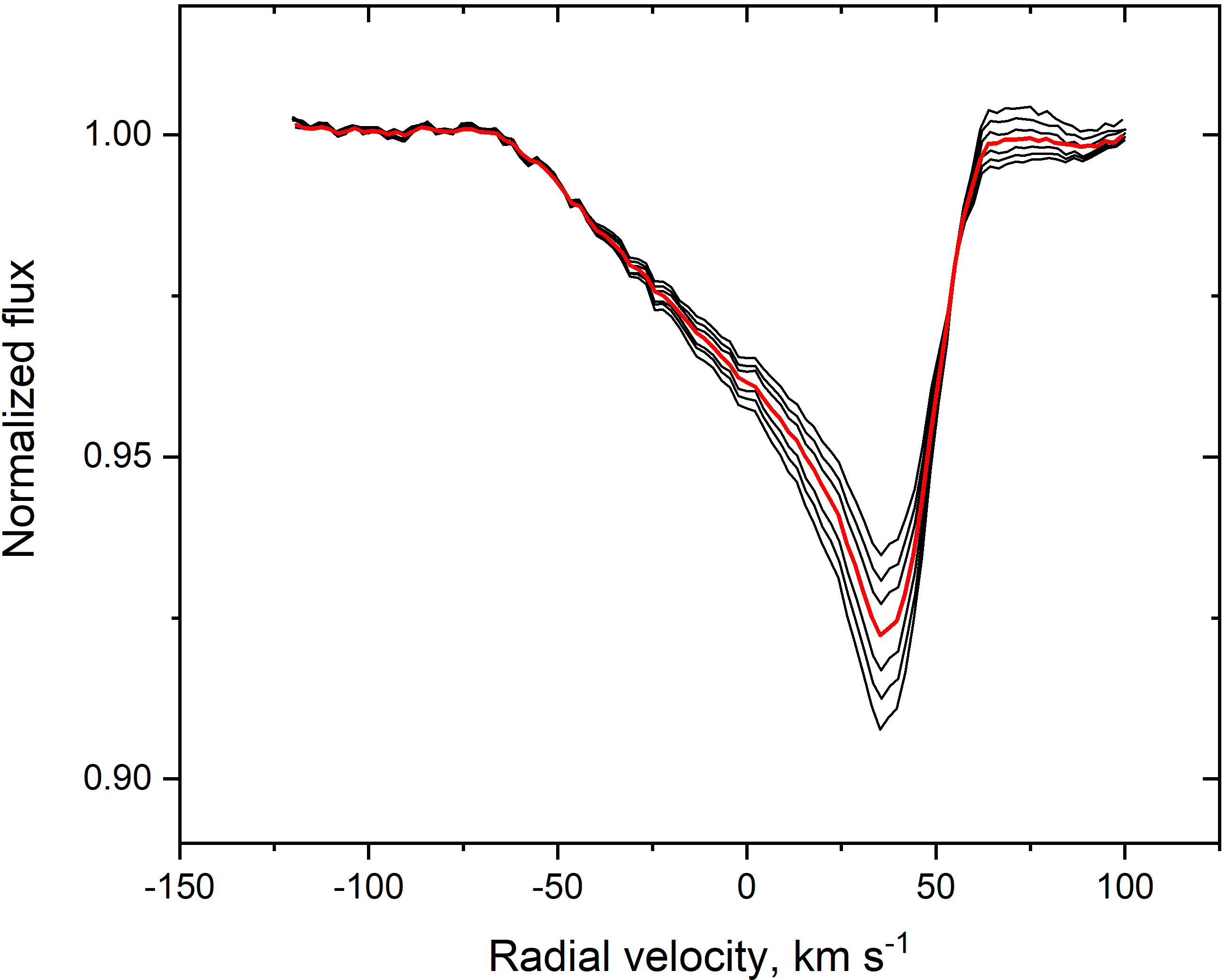}\hspace{5mm}
    \includegraphics[width=8.5cm]{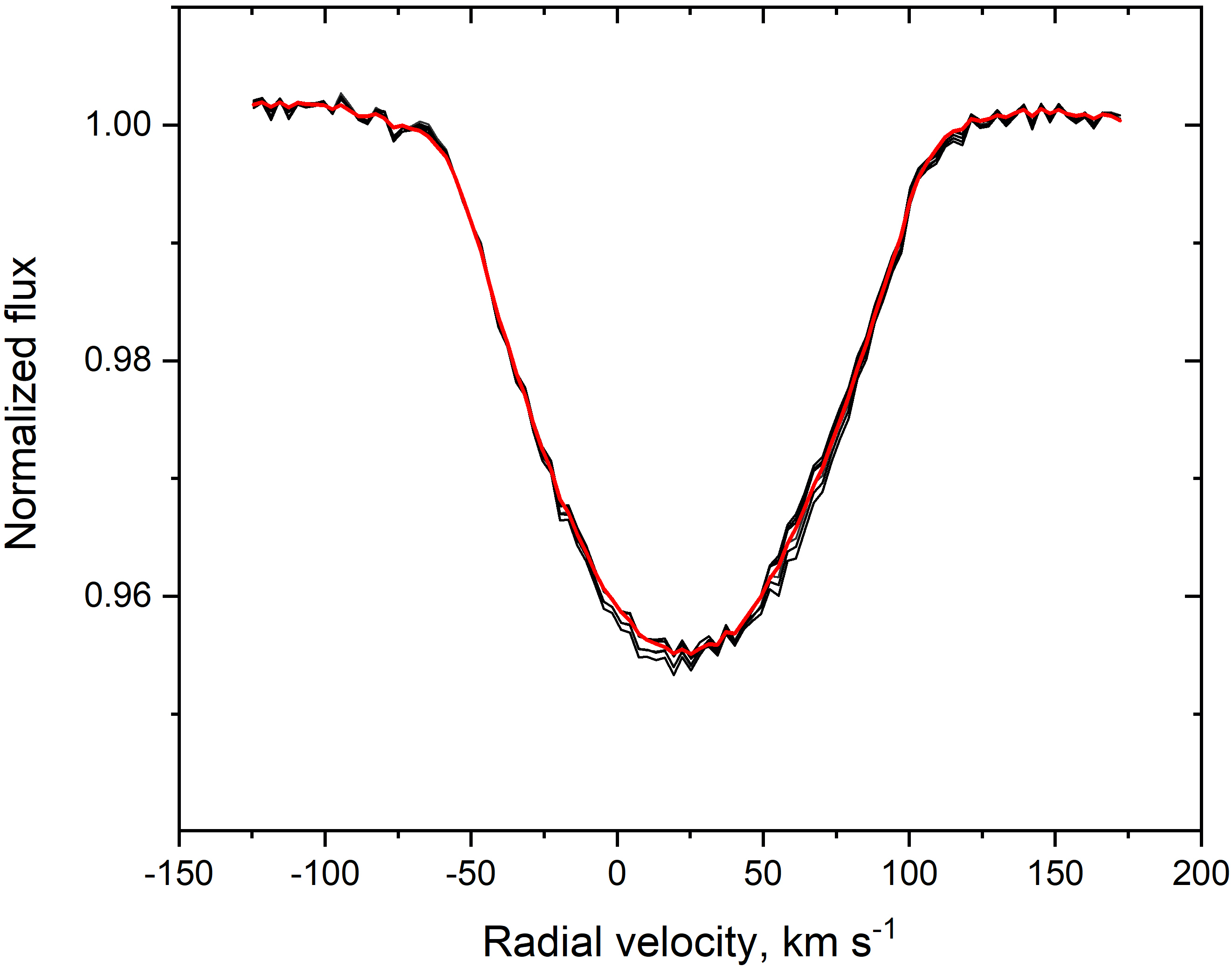}
      \caption{Effect of changing atmospheric parameters of the primary component of the RZ\,Cas system on the LSD profiles of the primary (left) and secondary (right). {\bf From top to bottom:} \teff, \logg, and [M/H] are varied between 8\,400 and 9\,200~K (step width of 100~K), 3.9 and 4.7~dex (step width of 0.1~dex), and -0.3 and 0.3~dex (step width of 0.1~dex), respectively. The LSD profiles shown in red correspond to the true parameters of the star listed in Table~\ref{Tab:RZCas_KIC11285625}.  
              }
         \label{Fig:RZCas_LSD_AtmosphericParametersChangePrimary}
   \end{figure*}

\begin{figure*}[h!]
   \centering
    \includegraphics[width=8.5cm]{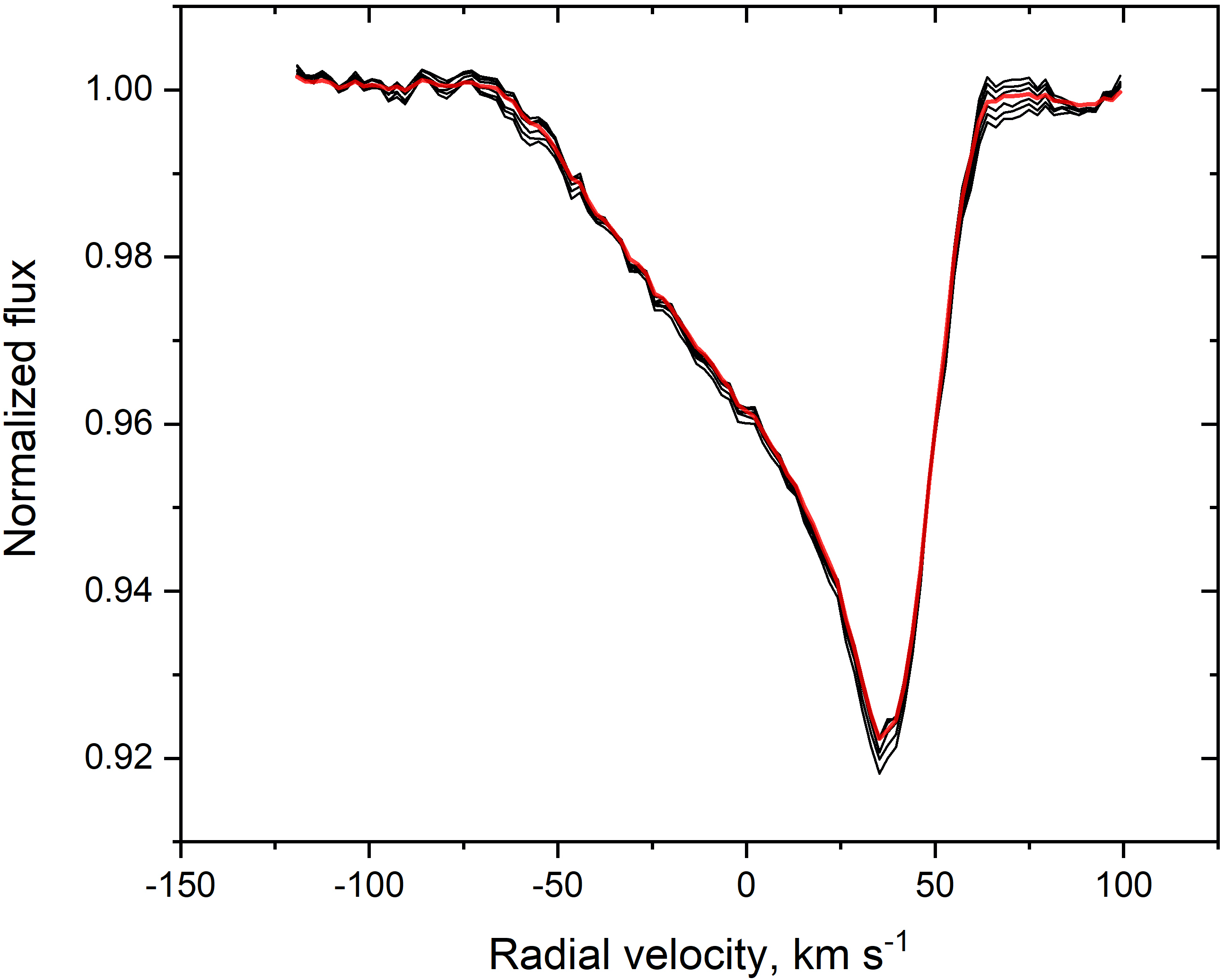}\hspace{5mm}
    \includegraphics[width=8.5cm]{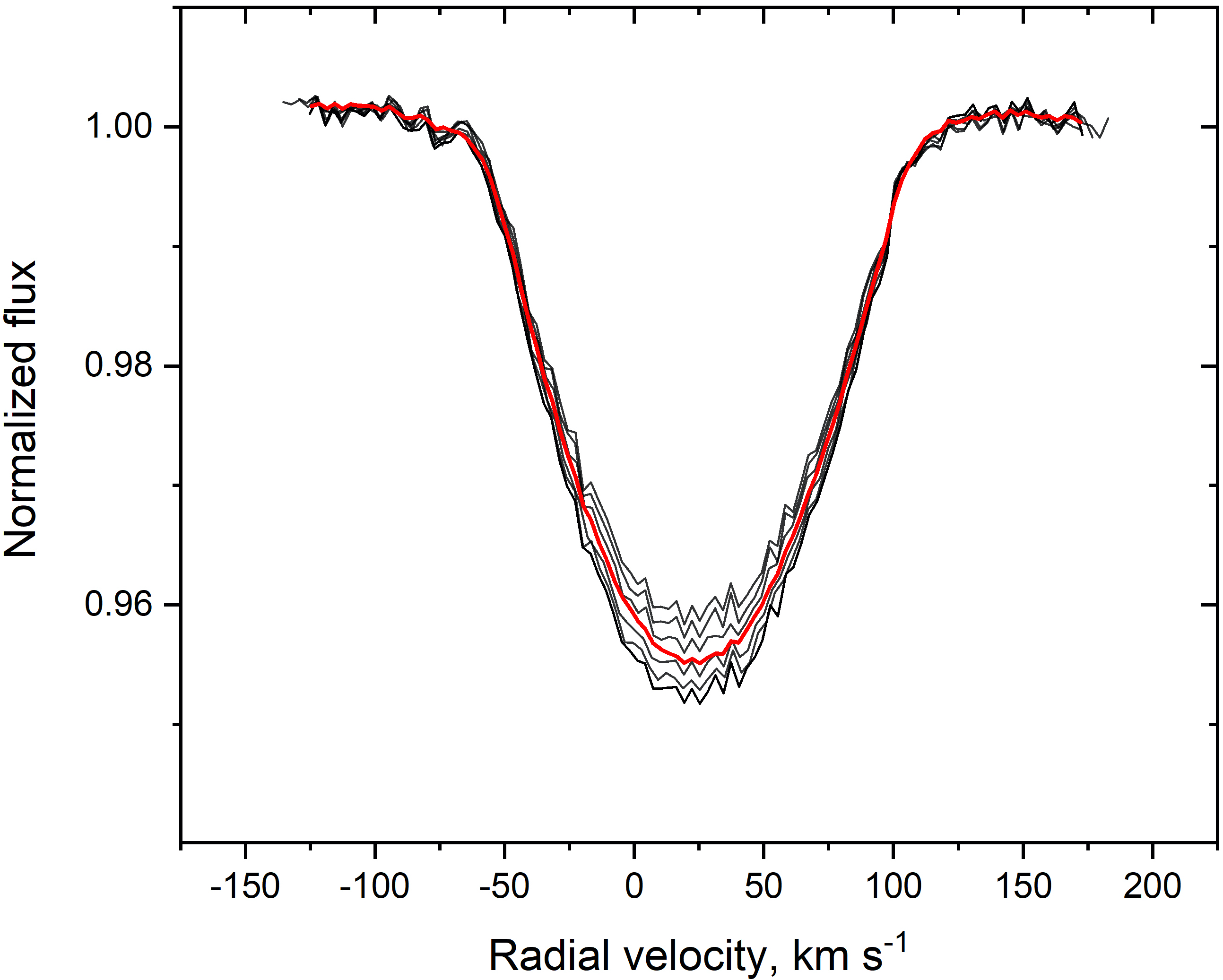}\vspace{5mm}
    \includegraphics[width=8.5cm]{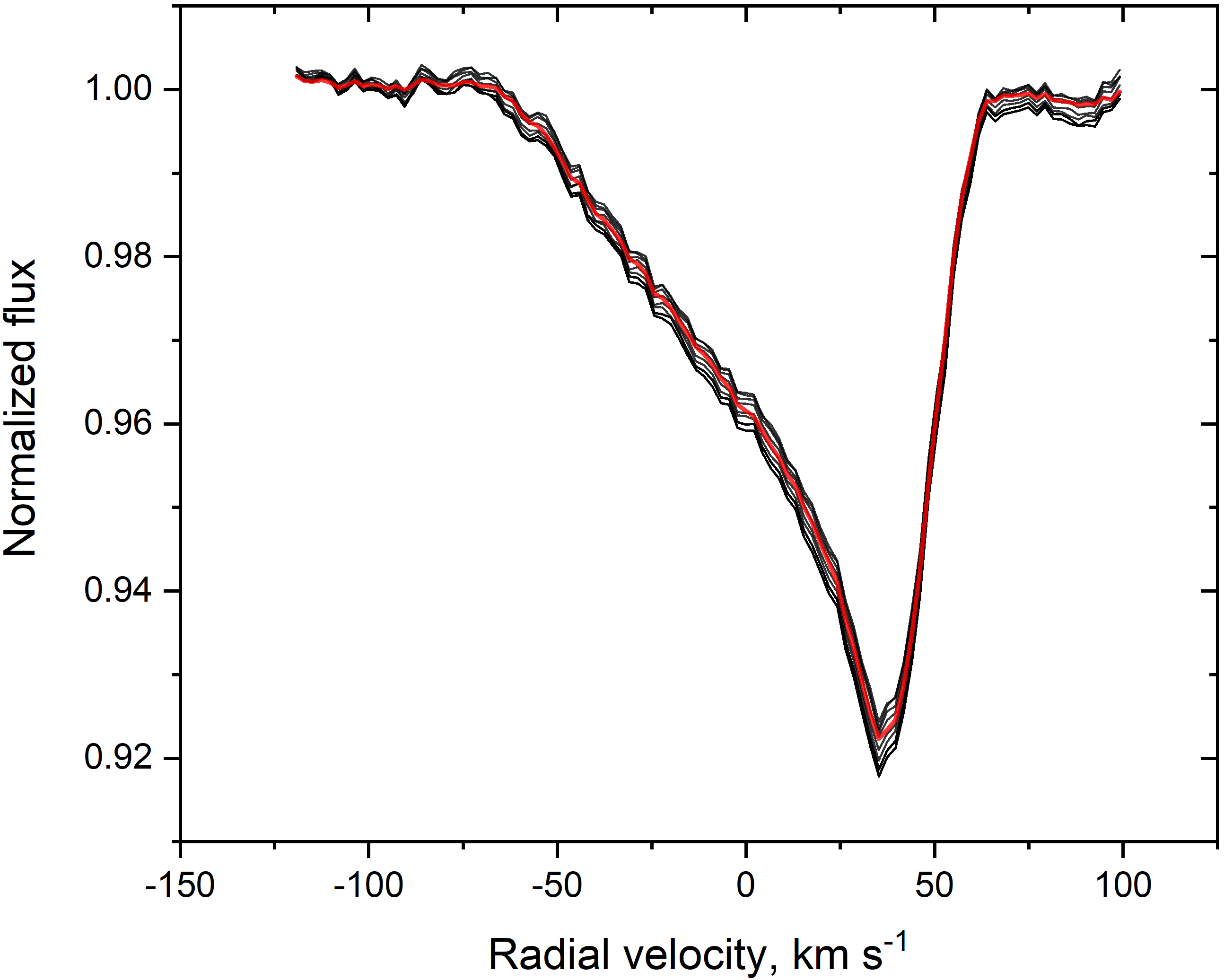}\hspace{5mm}
    \includegraphics[width=8.5cm]{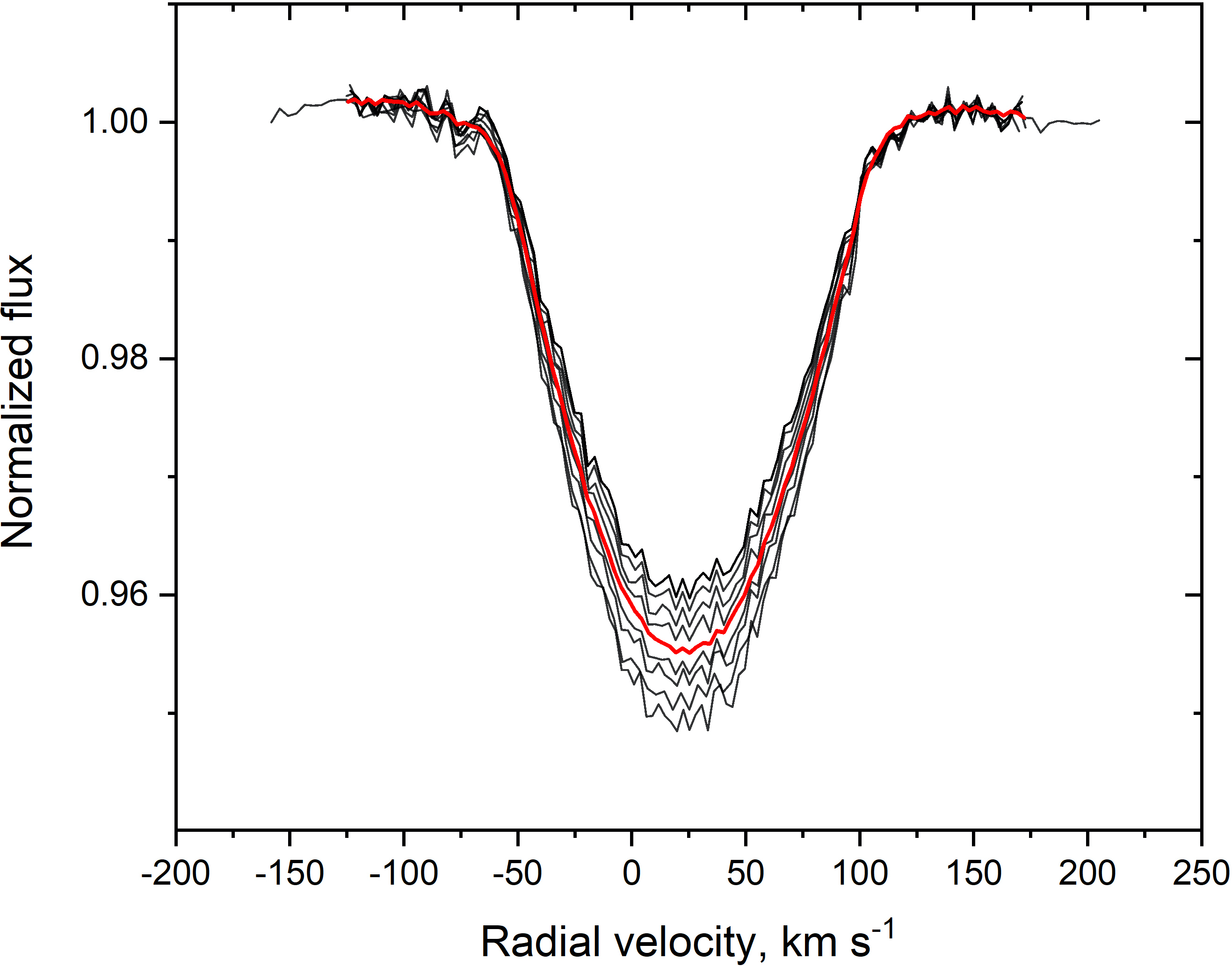}\vspace{5mm}
    \includegraphics[width=8.5cm]{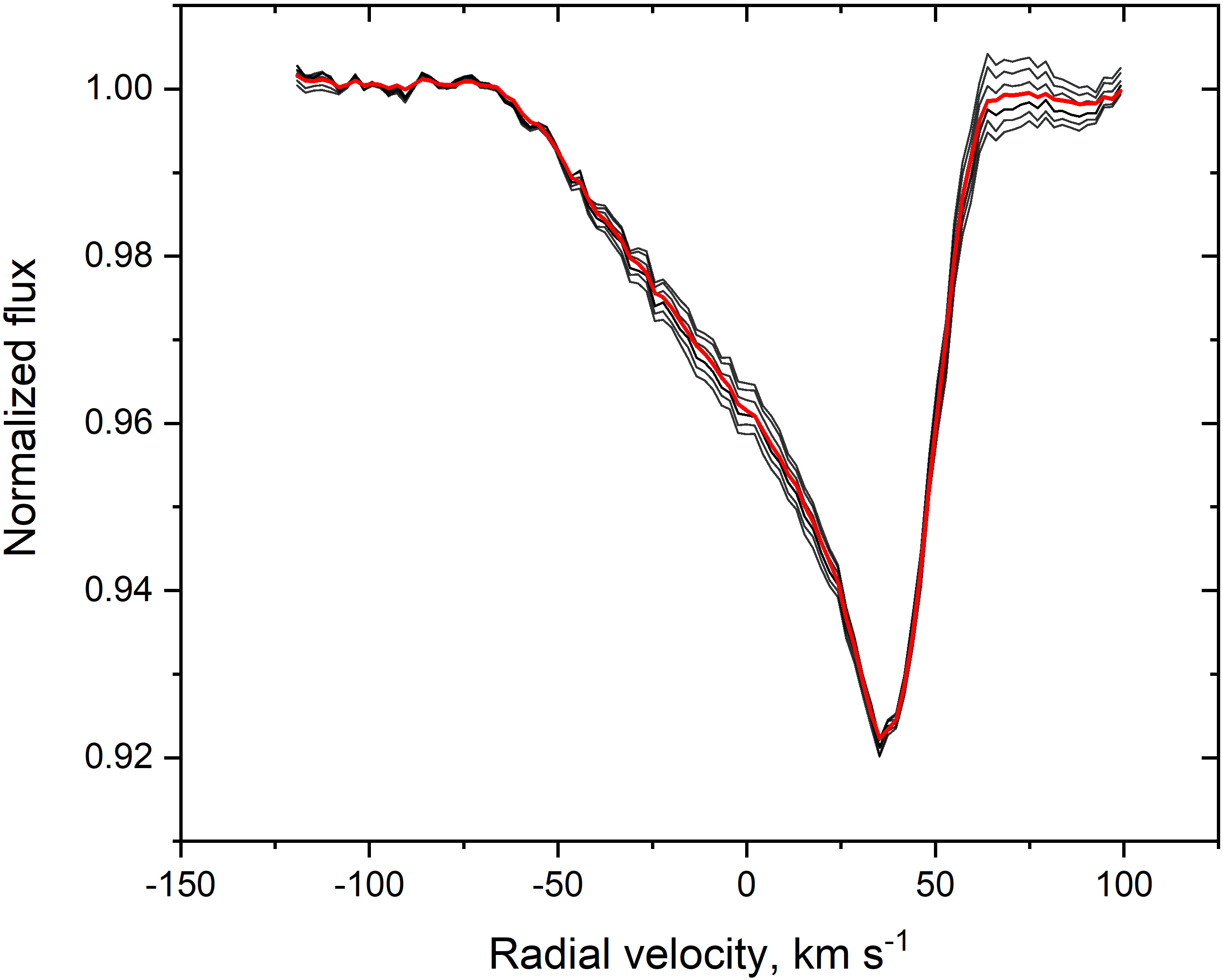}\hspace{5mm}
    \includegraphics[width=8.5cm]{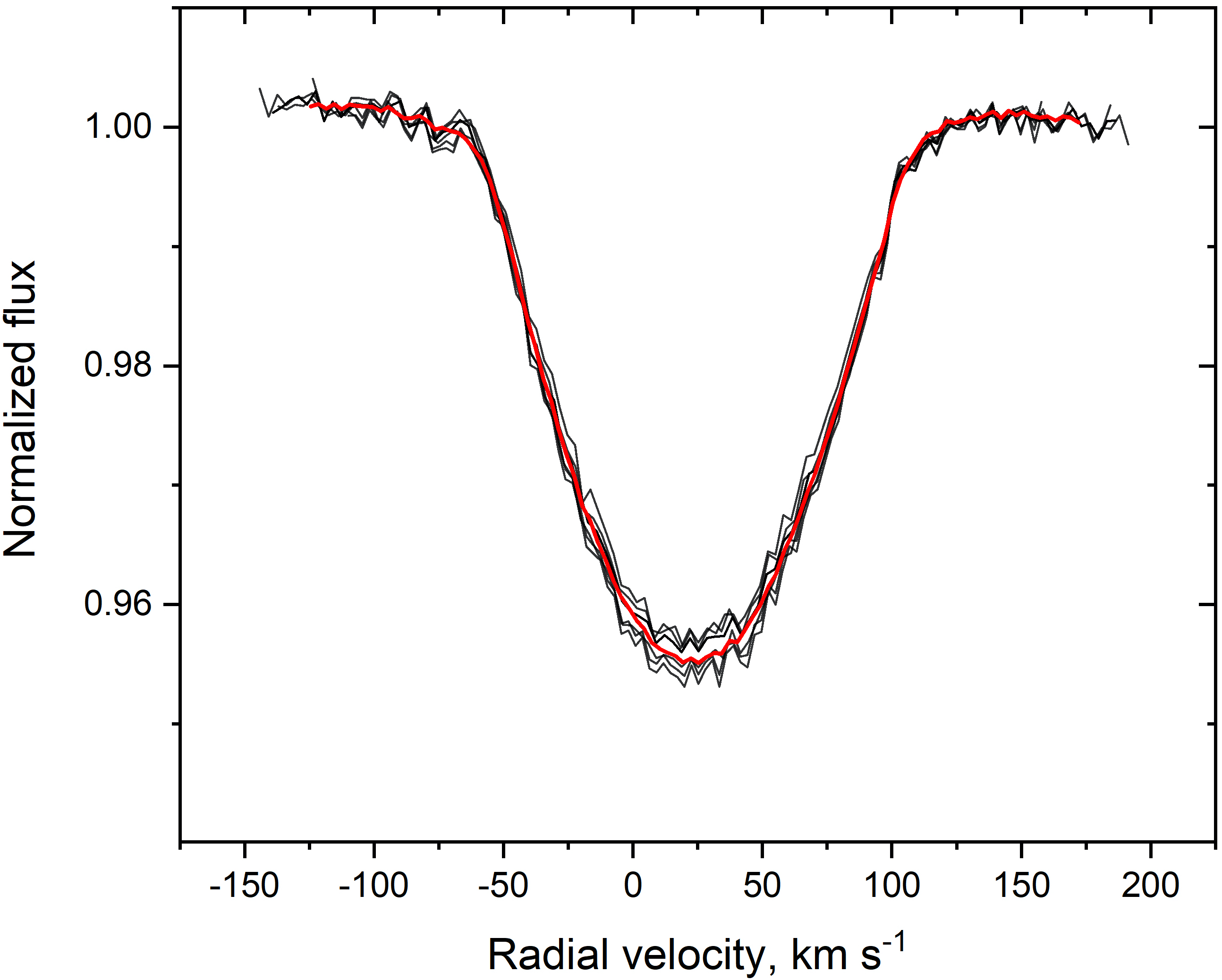}
      \caption{Same as Figure~\ref{Fig:RZCas_LSD_AtmosphericParametersChangePrimary} but for the effect of changing atmospheric parameters of the secondary component of the RZ\,Cas system. {\bf From top to bottom:} \teff, \logg, and [M/H] are varied between 4\,500 and 5\,100~K (step width of 100~K), 3.1 and 3.9~dex (step width of 0.1~dex), and -0.3 and 0.3~dex (step width of 0.1~dex), respectively. The LSD profiles shown in red correspond to the true parameters of the star listed in Table~\ref{Tab:RZCas_KIC11285625}.  
              }
         \label{Fig:RZCas_LSD_AtmosphericParametersChangeSecondary}
   \end{figure*}

In this Section, we present a few key tests based on simulated time-series of spectra, that are necessary to get a grasp on applicability range of the {\sc LSDBinary} algorithm. In these simulations, we consider two artificial binary systems representative of two particular classes of binary stars. Namely, detached systems with similar stellar components and semi-detached Algol-type systems with one component being significantly cooler and more evolved than the other. RZ\,Cas \citep[e.g.,][]{Soydugan2006,Tkachenko2009} and KIC\,11285625 \citep{Debosscher2013} are used as reference systems in our calculations, their relevant systemic and atmospheric parameters are summarised in Table~\ref{Tab:RZCas_KIC11285625}. The {\sc RME} module is used to synthesise composite spectra of both binary systems with a particular focus on the in-eclipse phases. These orbital phases are by far the most challenging because: (i) spectral line contributions of the two binary components overlap in the velocity space, and (ii) line profiles of the primary component are largely distorted due to the RM effect, and (iii) flux contribution of the primary component is orbital phase-dependent.

\subsection{Effects of the signal-to-noise ratio and resolving power}

In the first instance, we test the {\sc LSDBinary} method as to its ability to handle spectra of variable S/N and resolving power $R$. To that end, we add Poisson noise to our synthesised time-series to simulate spectra of S/N = 40, 80, and 120, for which we also consider three regimes of the resolving power: low, medium, and high at $R$ = 5\,000, 25\,000, and 60\,000, respectively. Figures~\ref{Fig:LSDBinary_RZCas_artificial} and \ref{Fig:LSDBinary_KIC11285625_artificial} summarise the results obtained for both binary systems. One can see that the common feature to both systems is the ability of the {\sc LSDBinary} algorithm to handle spectra in the entire range of S/N values: the obtained LSD profiles as well as LSD-based model spectra both have S/N that is (as expected) substantially higher than in the input spectra while the line profile shapes do not display any significant dependency on the S/N value. This is not an unexpected result given that the LSD technique is by design most efficient for stars that display large number of lines in their spectra, and both our simulated binary systems with their A- and F-type primary components meet the above requirement. Furthermore, we do not observe any significant changes in the obtained Level-0 data products when degrading resolving power from high ($R$ = 60\,000) to medium ($R$ = 25\,000) for the case of the RZ\,Cas system. However, we lose the ability to resolve the primary component's spectral line distortions due to the RM effect when degrading spectral resolution further to its lowest value of $R$ = 5\,000. Comparing the cases of RZ\,Cas and KIC\,11285625 with each other, we find that the effect of spectral resolution is much more important in the latter case: local distortions of the LSD profile of the primary component in the KIC\,11285625 system get smeared out at medium resolving power of $R$ = 25\,000 already. This is likely explained by the much smaller value of the projected rotational velocity of the star than in the case of the primary component of RZ\,Cas and, to a lesser extent, by relative contributions of the two stellar components of KIC\,11285625 to its composite spectrum. Overall, orbital and stellar configuration of the KIC\,11285625 system suggests much lower amplitude of the RM effect than in the case of RZ\,Cas, which in turn requires higher spectral resolution to resolve the effect. 

\begin{figure*}
   \centering
   \includegraphics[width=8.5cm]{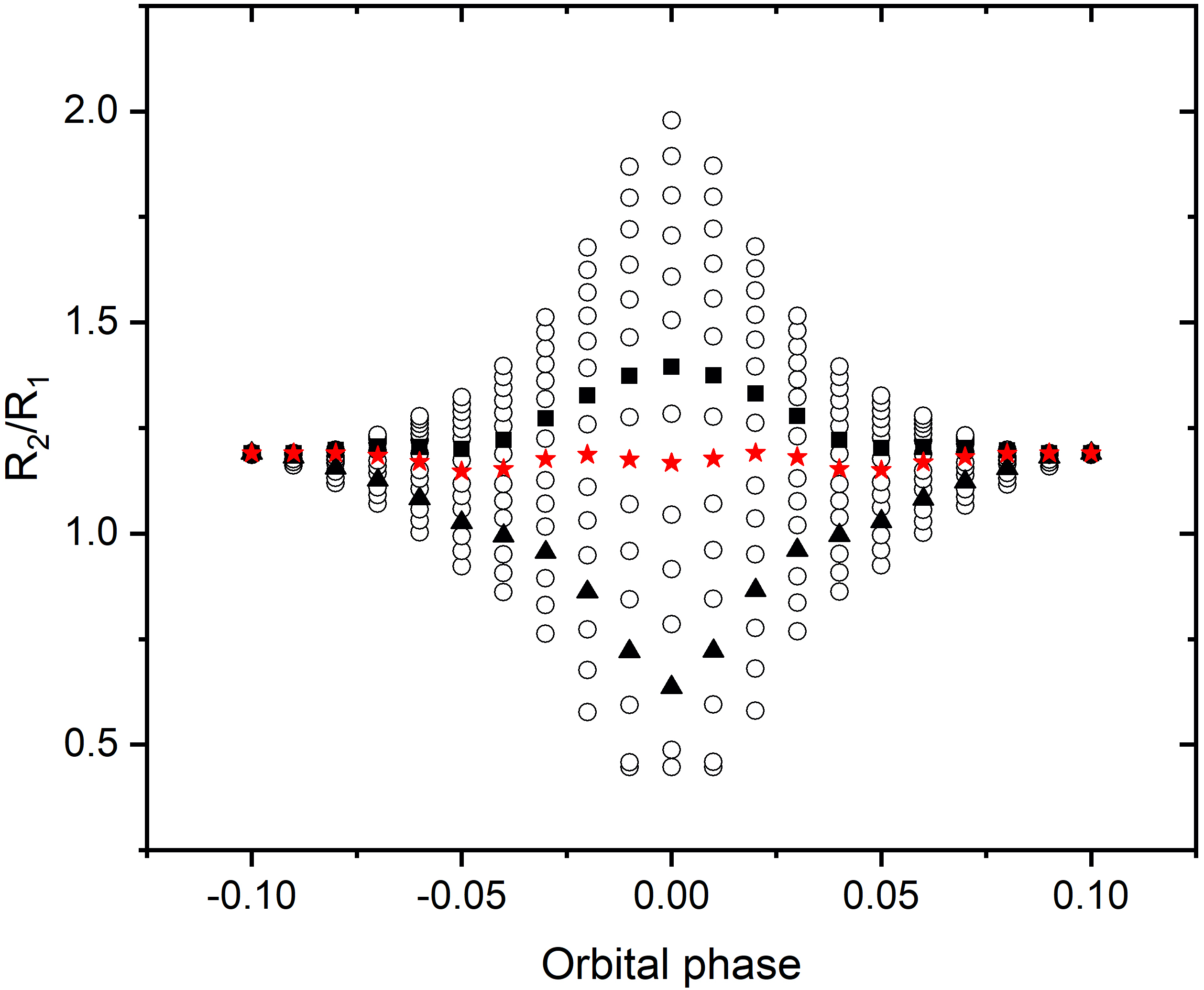}\hspace{5mm}
   \includegraphics[width=8.5cm]{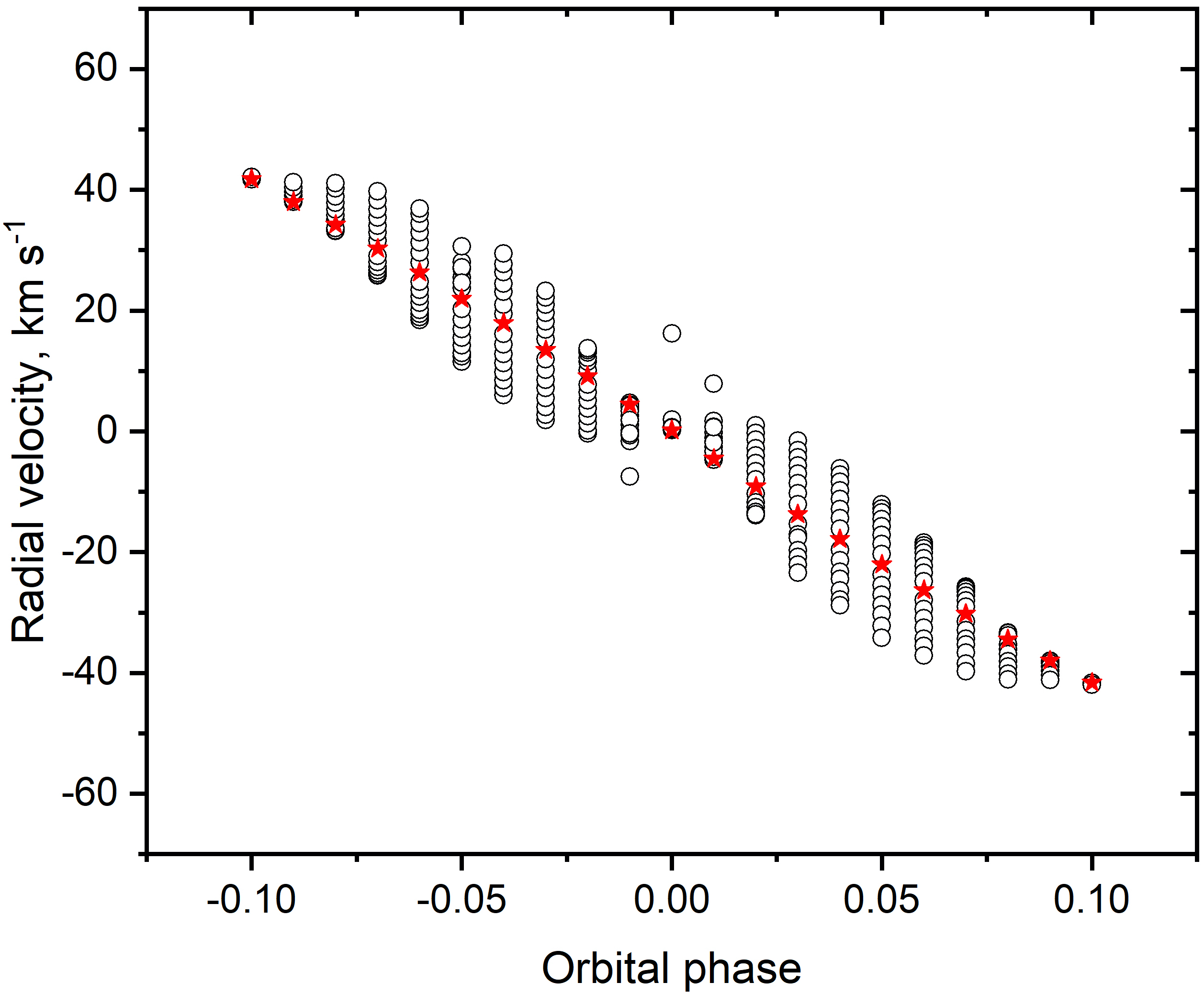}\vspace{5mm}
   \includegraphics[width=8.5cm]{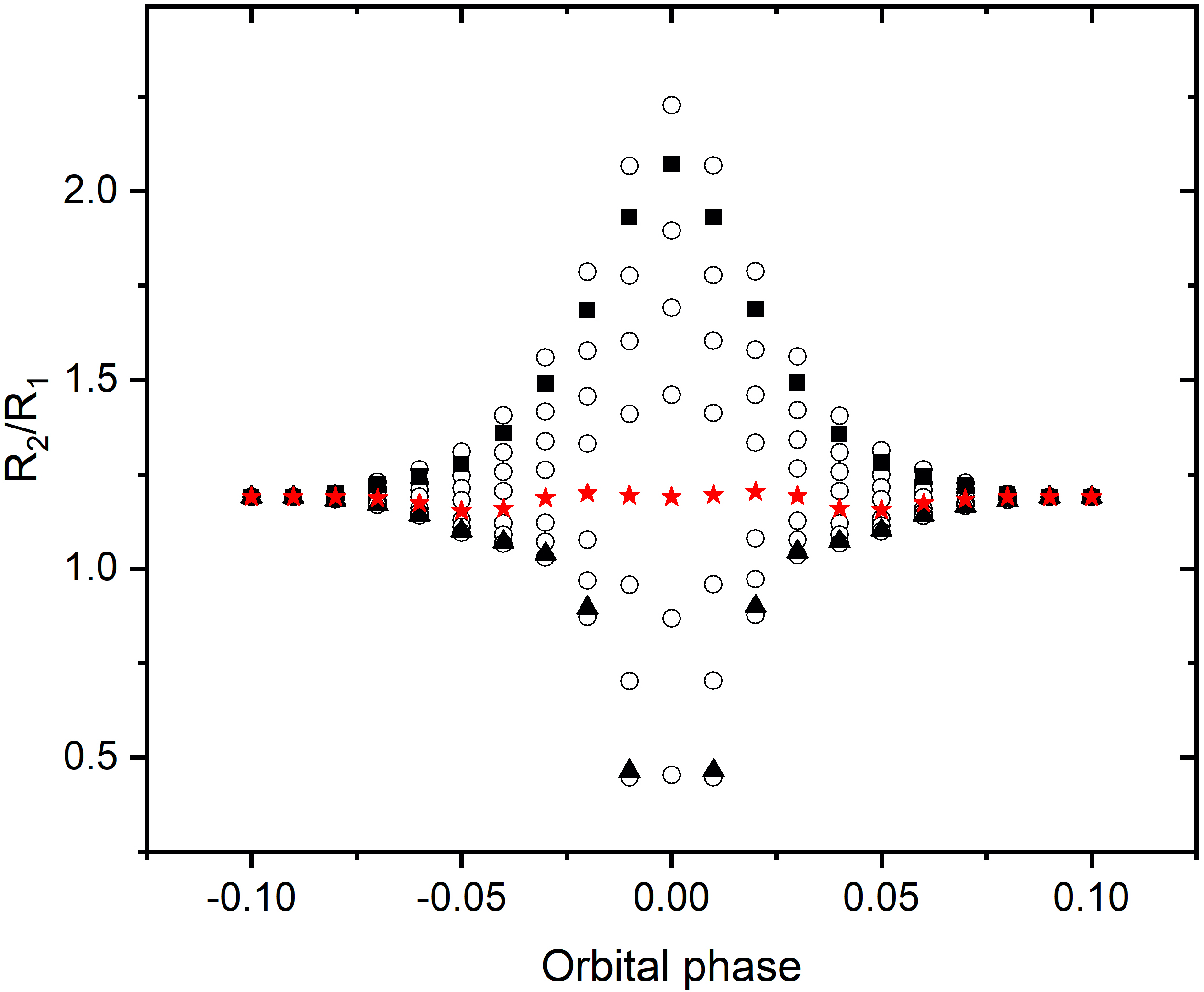}\hspace{5mm}
   \includegraphics[width=8.5cm]{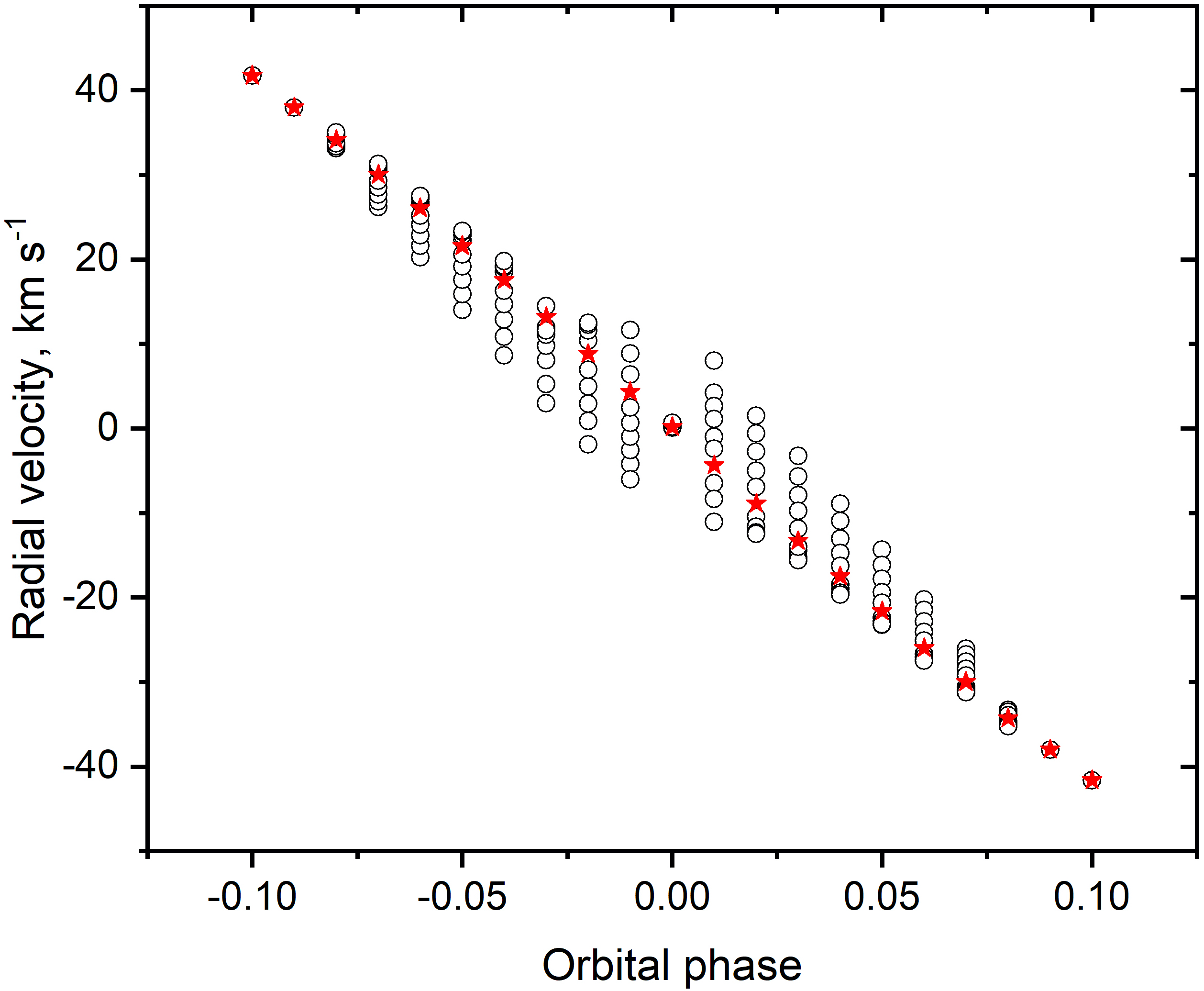}
      \caption{The effect of variable initial guess radii ratio $R_2/R_1$ (top row) and orbital inclination angle $i$ (bottom row) parameters on the obtained with the {\sc LSDBinary} algorithm orbital phase-resolved radii ratio (left column) and radial velocity (right column) curves. Filled squares and triangles in the top left panel refer to the initial guess $R_2/R_1$ = 1.1 and 1.4, respectively. The same symbols in the bottom left panel refer to the initial guess inclination angle $i$ values of 74$^{\circ}$ and 88$^{\circ}$, respectively. Red stars indicate a solution corresponding to the parameter values of $R_2/R_1$=1.2 and $i$=82$^{\circ}$ - a combination that represents the closest match to the true parameter values listed in Table~\ref{Tab:RZCas_KIC11285625}. See text for details.}
         \label{Fig:R2R1_InclinationAngle_Change_RME}
   \end{figure*} 

\subsection{Effect of variable input atmospheric parameters}
\label{Sect:AtmosphericParametersEffect}
In the next test, we investigate the effect of changing atmospheric parameters of either binary component on the Level-0 data products of the {\sc LSDBinary} algorithm. The effect of variable/uncertain atmospheric parameters is expected to propagate into the LSD profiles through changing properties of the input line mask as well as through the associated change in the components' flux ratio. Since the effect is qualitatively similar for both simulated binary systems, here we present the results obtained for the RZ\,Cas system only. We also present results for a single combination of $R$ = 60\,000 and S/N = 120 since the effect of changing these characteristics has been discussed in the previous section.

Figures~\ref{Fig:RZCas_LSD_AtmosphericParametersChangePrimary} and \ref{Fig:RZCas_LSD_AtmosphericParametersChangeSecondary} illustrate the effect of changing atmospheric parameters of the primary and secondary components, respectively, on the obtained LSD profiles. Effective temperature $T_{\rm eff}$, surface gravity $\log\,g$, and metallicity [M/H] are varied symmetrically with respect to their true values (see Table~\ref{Tab:RZCas_KIC11285625}) and within typical spectroscopic uncertainties: $\sim$5\% in \teff\ and 0.3-0.4~dex in \logg\ and [M/H]. The LSD profile of the primary changes most notably when varying the component's \teff\ and [M/H] (top and bottom left panels in Figure~\ref{Fig:RZCas_LSD_AtmosphericParametersChangePrimary}) while it appears to be barely sensitive to the variations of $\log\,g$ (middle left panel in Figure~\ref{Fig:RZCas_LSD_AtmosphericParametersChangePrimary}). Lack of sensitivity to the latter parameter is likely explained by limited effect of the surface gravity of the star on the overall shape and, to a lesser extent, strength of metal lines in the \teff\ regime of A- and F-type stars. At the same time, changing parameters of the primary component have little to no effect on the LSD profile of the secondary star (right column in Figure~\ref{Fig:RZCas_LSD_AtmosphericParametersChangePrimary}). 

Similarly, varying parameters of the secondary star has little to no effect on the LSD profile of the primary (left column in Figure~\ref{Fig:RZCas_LSD_AtmosphericParametersChangeSecondary}). At the same time, the LSD profiles of the secondary component appear to be most sensitive to changes in $T_{\rm eff}$ (top right panel in Figure~\ref{Fig:RZCas_LSD_AtmosphericParametersChangeSecondary}) and $\log\,g$ (middle right panel in Figure~\ref{Fig:RZCas_LSD_AtmosphericParametersChangeSecondary}) of the star, and less so to its metallicity [M/H] (bottom right panel in Figure~\ref{Fig:RZCas_LSD_AtmosphericParametersChangeSecondary}). We note that the secondary component in an Algol-type system is a more evolved and about twice as cool as the primary which explains a more appreciable sensitivity of its LSD profile to the \logg\ parameter. 

Ultimately, we stress that no significant changes in the overall shape of the LSD profile are observed for either of the binary components when their atmospheric parameters are varied within typical spectroscopic uncertainties. Instead, Figures~\ref{Fig:RZCas_LSD_AtmosphericParametersChangePrimary} and \ref{Fig:RZCas_LSD_AtmosphericParametersChangeSecondary} reveal an appreciable effect on the global scaling of the component's LSD profile in terms of its depth/strength. These findings are in line with the results by \citet{Tkachenko2013} who also demonstrate that ability of the LSD technique to reproduce an overall shape of the line profile is independent of the assumption of atmospheric parameters for the line mask calculation, provided they do not deviate from their true values by more than typical spectroscopic uncertainties.
 
\subsection{The effect of variable atmospheric parameters, radii ratio and orbital inclination}

In Section~\ref{Sect:Pseudo-code}, we briefly touched upon how the {\sc LSDBinary} algorithm can be used in conjunction with the assumption of a symmetric initial guess LSD profile to measure the in-eclipse RV variations associated with both the orbital motion of the star and the geometric Rossiter-McLaughlin effect. Such RV-curves can be used for follow-up studies, e.g. to assess the degree of spin-orbit misalignment, if present in the system. At the same time, we showed that the assumption of orbital phase-independent flux ratio of the two stars (that comes along with the assumption of a symmetric initial guess LSD profile) results in an appreciably variable with orbital phase radii ratio $R_2/R_1$ delivered by {\sc LSDBinary}. Obviously, true radii of the stars (and hence their ratio) do not change in the course of the eclipse and the observed variation of $R_2/R_1$ is a pure effect of ignoring orbital phase dependency of the flux ratio in the initial guess calculations. In essence, the $R_2/R_1$ parameter is being exploited in this case as a global scaling factor for the light contribution of the primary component that varies in the course of its eclipse.

\begin{figure*}
   \centering
   \includegraphics[width=8.5cm]{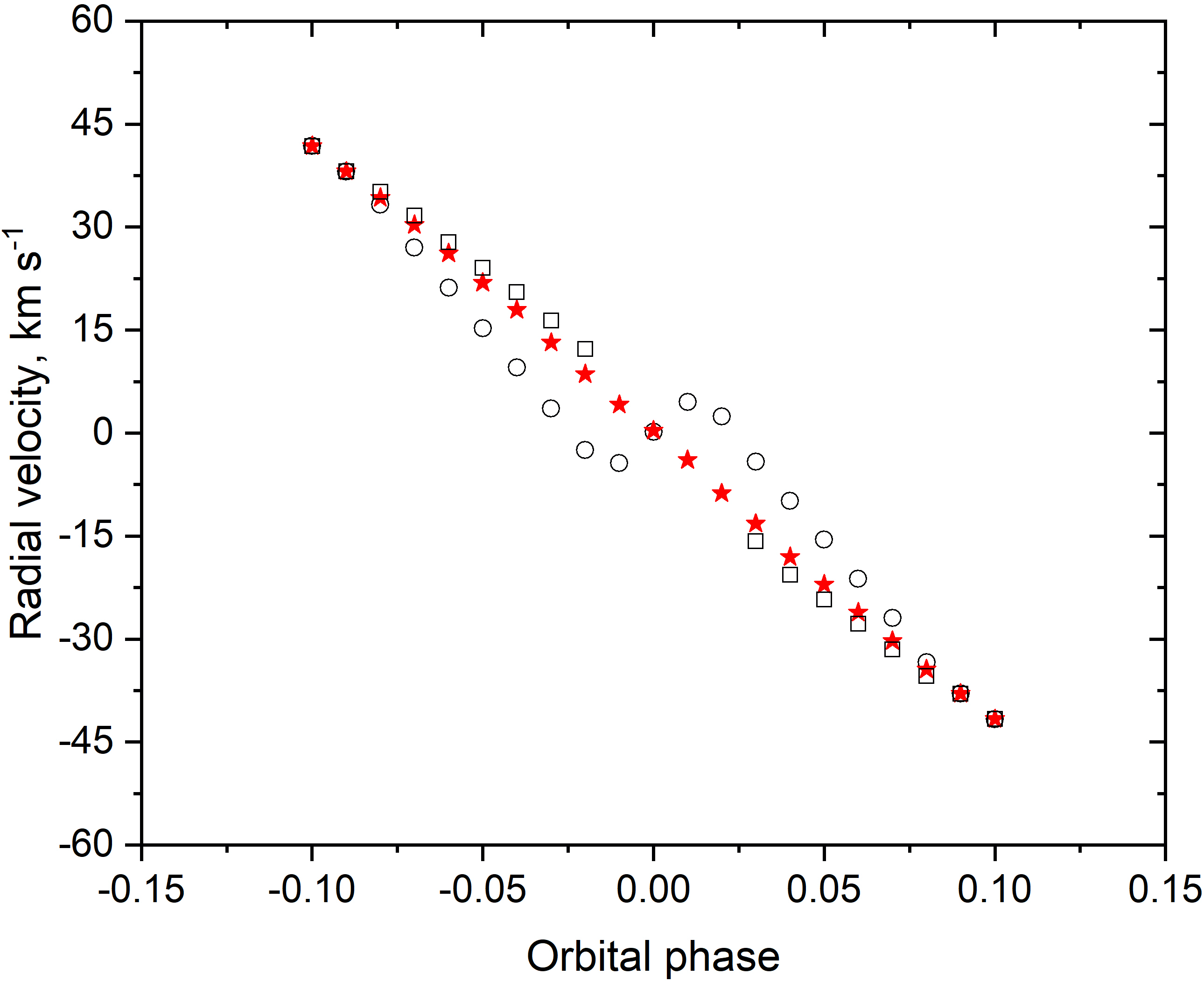}\hspace{5mm}
   \includegraphics[width=8.5cm]{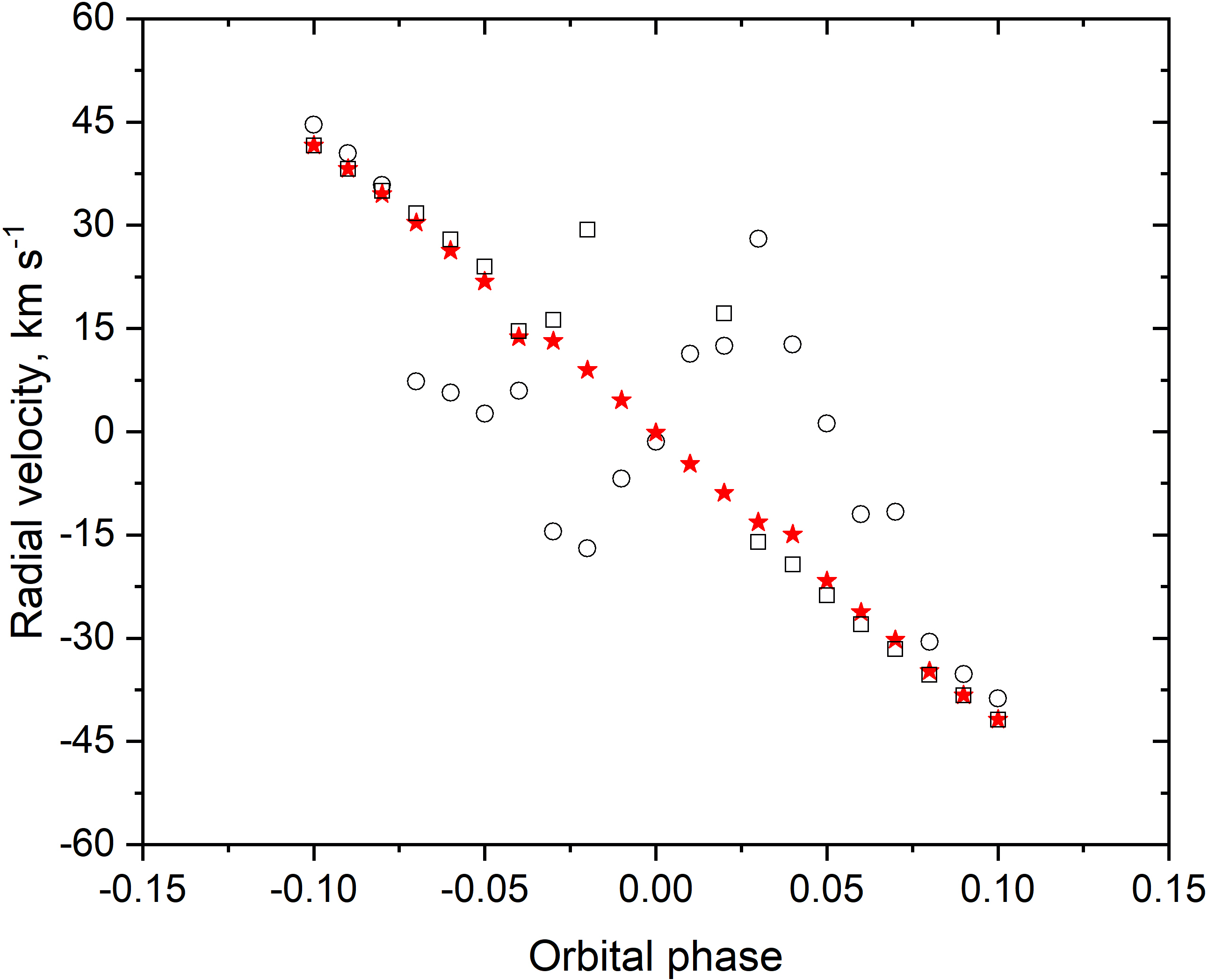}\vspace{5mm}
   \includegraphics[width=8.5cm]{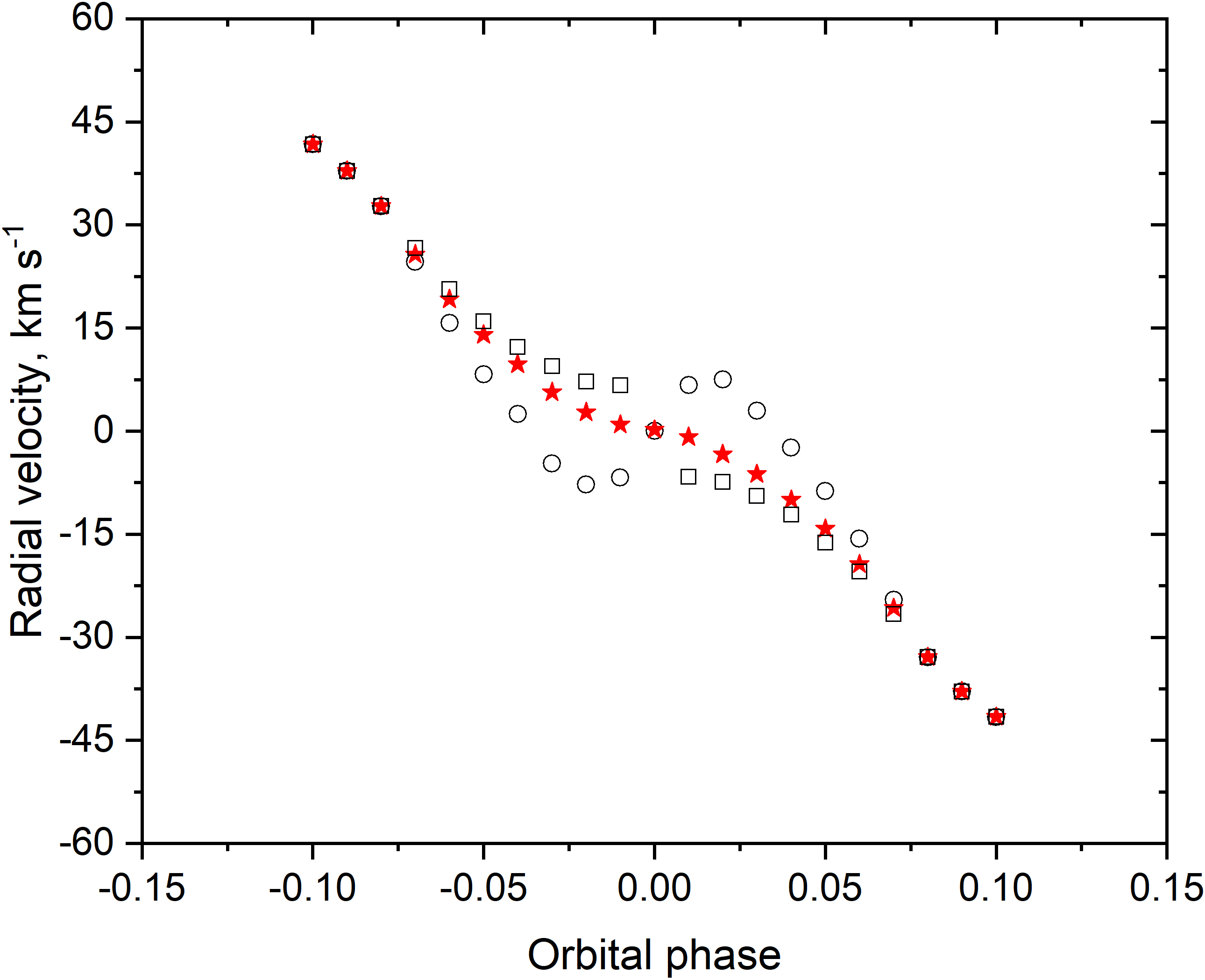}\hspace{5mm}
   \includegraphics[width=8.5cm]{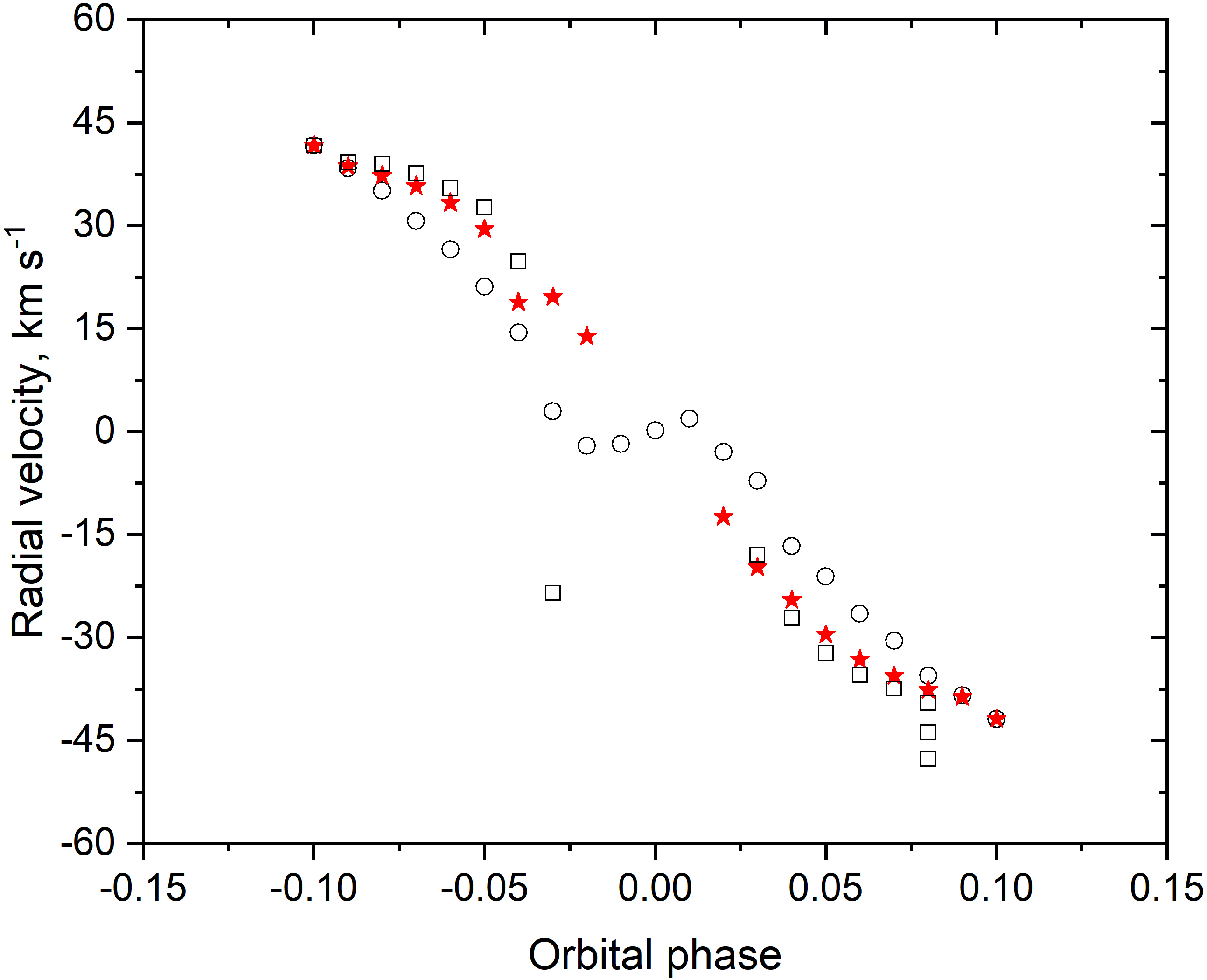}
      \caption{{\bf Top row:} the effect of incorrectly assumed initial guess $T_{\rm eff}$ of the primary component with $R_2/R_1$ being fixed to its true value of 1.19 (see Table~\ref{Tab:RZCas_KIC11285625}). Left and right panel shows the effect of under- and over-estimated effective temperature value, i.e. \teff=8\,500~K and 9\,100~K, respectively. Red stars show the optimal solution obtained with {\sc LSDBinary} corresponding to the orbital inclination angle $i=82^{\circ}$. Two more solutions corresponding to $i=74^{\circ}$ (open circles) and 90$^{\circ}$ (open squares) are shown for comparison. {\bf Bottom row:} the combined effect of incorrectly assumed initial guess \teff\ of the primary and components' radii ratio $R_2/R_1$. Left and right panel shows the case of under- and over-estimated parameters: \teff = 8\,500~K and $R_2/R_1$ = 1.02, and \teff = 9\,100~K and $R_2/R_1$ = 1.36, respectively.  See text for details.}
         \label{Fig:InclinationAngleOptimisation_TeffUncertainty}
   \end{figure*} 

In this section, we investigate the effect of variable (at the step of the initial guess calculation in the {\sc LSDInit} module) stellar radii ratio and orbital inclination angle on the data products of the {\sc LSDBinary} algorithm. Similar to Section~\ref{Sect:AtmosphericParametersEffect}, we focus on the case of an artificial Algol-type system resembling stellar and orbital properties of the RZ~Cas binary system (see Table~\ref{Tab:RZCas_KIC11285625}), where the artificial ``observed'' spectra are characterised by $R = 60\,000$ and S/N = 120. To account for the Rossiter-McLaughlin effect in the test, the initial guess orbital phase-resolved LSD profiles and wavelength-dependent functional form of the components' flux ratio are computed with the {\sc RME} module.

In the current test, the radii ratio $R_2/R_1$ and orbital inclination angle $i$ parameters are varied one at the time while keeping the other one fixed. This way, the two scenarios considered are: (i) the $R_2/R_1$ parameter is varied in the range from 0.8 to 1.6 (step width of 0.05) while keeping $i$ = 82$^{\circ}$, and (ii) the $i$ parameter is varied in the range from 70$^{\circ}$ to 90$^{\circ}$ (step width of 2$^{\circ}$) while keeping $R_2/R_1$ = 1.19. In total, we consider 28 $R_2/R_1$-$i$ parameter combinations (17 and 11 in the cases (i) and (ii), respectively) when computing the initial guess LSD profiles and components' flux ratio with the {\sc LSDInit} module. The initial guess is then passed on to the {\sc LSDBinary} algorithm that takes the above-described artificial ``observed'' spectra as input and returns optimised LSD profiles of both binary components and their radii ratio parameter $R_2/R_1$. 

The top row in Figure~\ref{Fig:R2R1_InclinationAngle_Change_RME} summarises the results obtained for the case (i) where the initial guess radii ratio parameter $R_2/R_1$ is varied while keeping the inclination angle $i$ fixed. The left and right panels demonstrate the orbital phase-resolved $R_2/R_1$ and RV curves, respectively, where the parameter combination that represents the closest match to the true parameter values is shown with red stars. One can see that this particular solution favours a constant with the orbital phase value of the radii ratio and the RV-curve that closely resembles pure orbital motion of the primary component. At the same time, setting the initial guess $R_2/R_1$ to a larger or smaller parameter value (e.g., 1.4 and 1.1 as indicated with the filled triangles and squares in the top left panel in Figure~\ref{Fig:R2R1_InclinationAngle_Change_RME}) results in the $R_2/R_1$ versus orbital phase curve that deviates appreciably from the constant value of the respective parameter. The same conclusion holds true for the RV-curve where one starts to see residual signal from the Rossiter-McLaughlin effect which increases progressively as the deviations of the initial guess $R_2/R_1$ parameter get larger from its true value. The former is the direct consequence of incorrectly assumed initial guess component's flux ratio due to variable $R_2/R_1$ and fixed effective temperatures, while the latter is explained by the variations in the obtained with {\sc LSDBinary} LSD profiles associated with the changing Rossiter-McLaughlin effect geometry due to variations in $R_2/R_1$.

Similar effect is observed when the initial guess orbital inclination angle parameter $i$ is varied while keeping $R_2/R_1$ fixed (see the bottom row in Figure~\ref{Fig:R2R1_InclinationAngle_Change_RME}). Changing the orbital inclination angle $i$ to a larger or smaller value (e.g., 88$^{\circ}$ and 74$^{\circ}$ as indicated in the bottom left panel with the filled triangles and squares, respectively) leads to a smaller or larger area of the primary component being visible during its eclipse phases. This in turn leads to a larger or smaller flux contribution from the primary star in the initial guess, the effect that is compensated by a larger or smaller value of the $R_2/R_1$ parameter as optimised in the {\sc LSDBinary} module. As can be seen in Figure~\ref{Fig:R2R1_InclinationAngle_Change_RME} (bottom left panel), the effect is progressive with orbital phase and reaches its point of extremum at the centre of the primary eclipse. Just as is the case with the variable initial guess $R_2/R_1$ parameter, the orbital inclination angle $i$ has strong effect on the shape of the initial guess LSD profile which leads to the presence of residual Rossiter-McLaughlin effect signal in the RV curve when the value of $i$ is not optimal (see bottom right panel in Figure~\ref{Fig:R2R1_InclinationAngle_Change_RME}).

In the last set of tests on artificial data, we investigate how the in-eclipse orbital-phase resolved RV curve of the eclipsed primary component is affected by uncertainty in the determination of its \teff\ and component's radii ratio $R_2/R_1$. To start with, we assume the $R_2/R_1$ parameter is known by other means, e.g. from a light curve solution of the system. In this particular scenario, we fix $R_2/R_1$ to its true value of 1.19 (see Table~\ref{Tab:RZCas_KIC11285625}, column ``RZ Cas'') and investigate how well we can recover the orbital inclination angle $i$ of the system when \teff\ of the primary is varied within its typical spectroscopic uncertainties of 3-5\%, or $\sim$300~K in the absolute value for the system considered here. Because the Rossiter-McLaughlin effect is accounted for in the calculation of the initial guess LSD profiles and components' flux ratio, the optimal value of the inclination angle is the one that provides minimal deviation of the RV curve obtained with the {\sc LSDBinary} algorithm from the component's RV curve that resembles its pure orbital motion. The top row in Figure~\ref{Fig:InclinationAngleOptimisation_TeffUncertainty} summarises the obtained results by considering two extreme values of the effective temperature of the primary component: 8\,500~K (left panel) and 9\,100~K (right panel). The results are illustrated for three values of the orbital inclination angle $i$: two extreme cases of 74$^\circ$ and 90$^{\circ}$, and the optimum value of 82$^{\circ}$ we arrived at. We do not employ any minimisation algorithm to search for an optimum value of the inclination angle, instead perform a grid search in the range i$\in$[74$^{\circ}$,90$^{\circ}$] with the step width of 1$^{\circ}$. One can see that true value of the inclination angle $i=82^{\circ}$ is recovered well in both cases considered in this test whereas other solutions are characterised by non-negligible deviations of individual RV measurements from those expected for pure orbital motion of the star. These results reinforce our conclusions made earlier in Section~\ref{Sect:AtmosphericParametersEffect} that any variations of atmospheric parameters that are within their typical spectroscopic uncertainties do not have significant effect on the shape of the obtained LSD profiles and hence RVs inferred from them.

Similarly, we now consider a case where the $R_2/R_1$ parameter is estimated with the {\sc LSDBinary} algorithm from the out-of-eclipse spectra instead of it being fixed to a priori known value. Considering the same two extreme values of \teff\ of the primary, i.e. 8\,500~K and 9\,100~K, we compute initial guess LSD profiles and component's flux ratio for the out-of-eclipse phases employing the {\sc SynthV} and {\sc Convolve} suite of codes. Therefore, we assume symmetric LSD profiles for both binary components and a constant flux ratio, a well-justified assumption for the out-of-eclipse phases. We then let {\sc LSDBinary} to optimise for the $R_2/R_1$ parameter by minimising the difference between the ``observed'' and LSD-based model composite spectra. We find $R_2/R_1 = 1.02$ and 1.36 to provide the best match between the model and ``observations'' when \teff\ of the primary is set to 8\,500~K and 9\,1000~K, respectively. The difference of some 15\% between the best fit $R_2/R_1$ and its true value is the direct consequence of incorrectly assumed \teff\ of the primary component. A lower (higher) initial guess value of the primary's effective temperature leads to over (under) estimation of its flux contribution to the total light of the system. The inconsistency arising between the model and ``observations'' is efficiently remediable by varying the $R_2/R_1$ parameter that, along with the effective temperatures, controls individual flux contributions of both binary components. Therefore, a lower (higher) initial guess \teff\ of the primary component is compensated by its larger (smaller) radius and leads to a decrease (increase) of the $R_2/R_1$ parameter. 

With the $R_2/R_1$ parameter being estimated from the out-of-eclipse-spectra, we now proceed with modelling of the in-eclipse spectra. The initial guess LSD profiles and component's flux ratio are computed with the {\sc RME} module for incorrectly assumed values of $R_2/R_1$ and \teff\ of the primary: (i) \teff = 8\,500~K, $R_2/R_1$ = 1.02, and (ii) \teff = 9\,100~K, $R_2/R_1$ = 1.36. The bottom row in Figure~\ref{Fig:InclinationAngleOptimisation_TeffUncertainty} shows the results obtained for the above cases (i) (left panel) and (ii) (right panel) and for three different values of the inclination angle $i$, 74$^{\circ}$ (open circles), 82$^{\circ}$ (red stars), and 90$^{\circ}$ (open squares). Although the RV curve corresponding to $i=82^{\circ}$ still provides the smallest deviations from the RV curve resembling pure orbital motion of the primary component compared to all other values of $i$ considered in this test, the residual distortions due to the Rossiter-McLaughlin effect are significant. In our previous test with variable initial guess \teff\ of the primary component (see above and top row in Figure~\ref{Fig:InclinationAngleOptimisation_TeffUncertainty}) we demonstrated that the effect of this parameter on the in-eclipse RV curve is minimal subject to the correct assumption of the orbital inclination angle. Therefore, we conclude that the $R_2/R_1$ parameter has a significantly larger effect on Level-0 and Level-1 data products of the {\sc LSDBinary} algorithm than effective temperature of the star, despite both of these parameters being important for a correct prediction of the binary component's flux ratio. This conclusion is also largely intuitive: whereas \teff\ of the primary impacts significantly its flux contribution to the total light of the system, the $R_2/R_1$ parameter also has a large effect on the geometry of the Rossiter-McLaughlin effect.

\section{Discussion and Conclusions}\label{Conclusions}

\begin{figure}
   \centering
   \includegraphics[width=8.5cm]{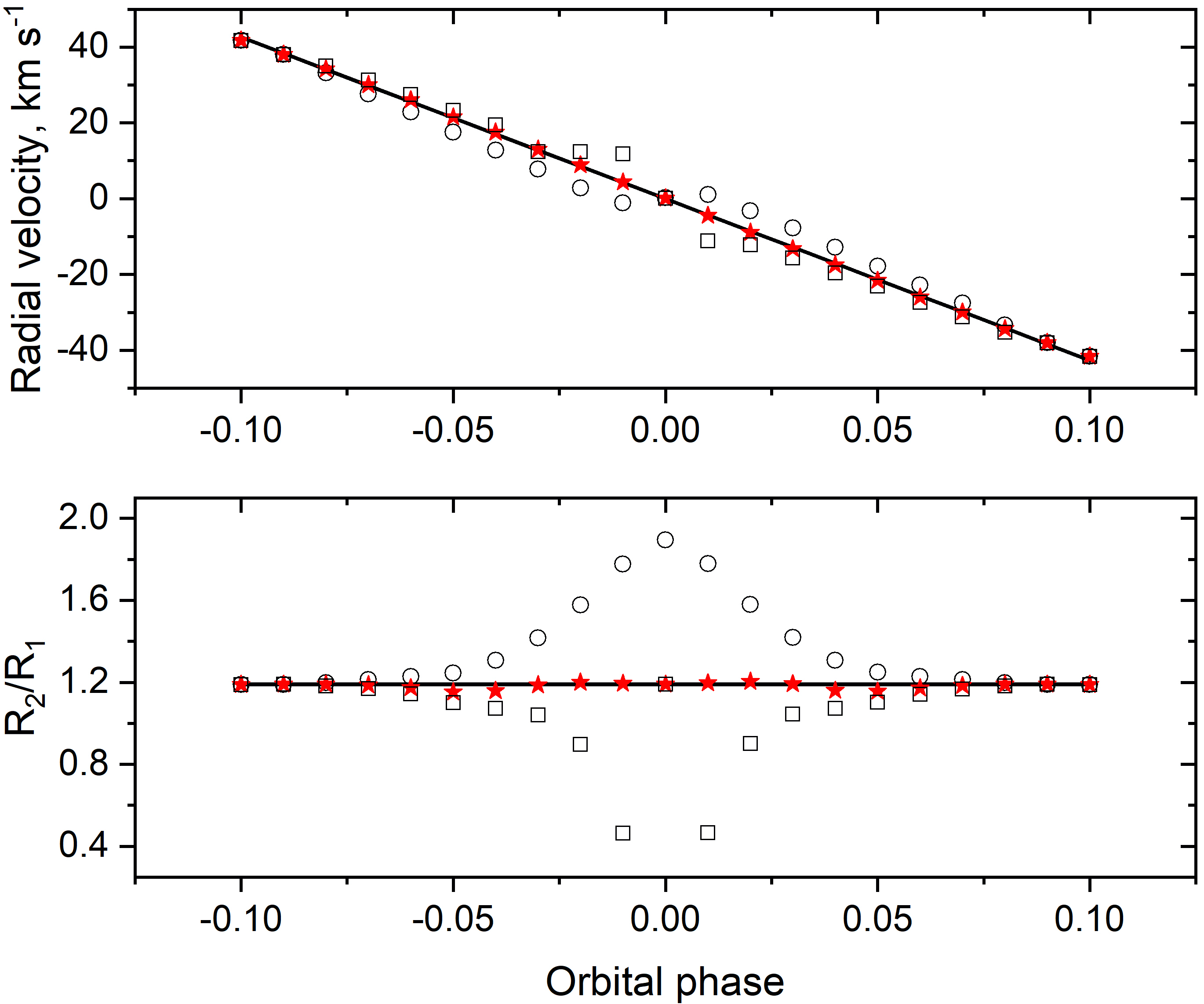}
      \caption{An example of fine-tuning the inclination angle parameter $i$ by minimising deviations of the $R_2/R_1$ (bottom) and RV (top) measurements obtained with the {\sc LSDBinary} algorithm from their respective constant value and pure orbital motion-driven curve. The open circles, red stars, and open squares show the results obtained for the orbital inclination angle $i$ = 76$^{\circ}$, 82$^{\circ}$, and 88$^{\circ}$, respectively. The black solid line shows the orbital motion-driven RV curve of the primary and the constant systemic value of $R_2/R_1=1.19$ in the top and bottom panel, respectively.}
         \label{Fig:InclinationAngleOptimisation}
   \end{figure}

In this study, we presented a generalisation of the Least-Squares Deconvolution technique to eclipsing, spectroscopic double-lined binary systems. The main focus is put on separation of spectral contributions of individual binary components in the velocity space so that their RVs can be measured from the obtained LSD profiles with high precision. In our tests, we put large emphasis on the in-eclipse phases where spectral contributions of the binary components are heavily blended. To that end, we develop a dedicated {\sc RME} code that allows one to compute initial guess LSD profiles and components' flux ratio in the presence of the Rossiter-McLaughlin effect. At the same time, we include an option of initial guess symmetric LSD profiles and a constant with orbital phase components' flux ratio to provide the means of working with the out-of-eclipse phases too. This latter option can also be applied to the in-eclipse orbital phases in which case a RV curve distorted due to the Rossiter-McLaughlin effect can be measured directly from the observations. 

The {\sc LSDBinary} software package developed in this work is freely available on GitHub\footnote{\url{https://github.com/AndrewStSp/LSDBinary}} and comprises: (i) the {\sc SynthV} and {\sc Convolve} suite of codes to compute line masks and initial guess symmetric LSD profiles; (ii) the {\sc RME} code to account for the Rossiter-McLaughlin effect in the calculation of the orbital phase-dependent initial guess LSD profiles and binary components' flux ratio; (iii) the {\sc LSDInit} module that serves as a wrapper around the above-mentioned codes and whose main purpose is to organise input required for the core calculations; and (iv) the {\sc LSDBinary} module that performs core calculations and, in its most basic configuration, delivers LSD profiles, LSD-based model spectra, and RVs of both binary components.

In this work, we also provide an extensive test of the {\sc LSDBinary} algorithm based on artificial spectra of two binary systems: an Algol-like system with a main-sequence A-type primary and evolved K-type giant secondary components, and a detached binary with two F-type components. The main results and conclusions of our study can be summarised as follows:

\begin{itemize}
    \item the algorithm has low sensitivity to signal-to-noise ratio of the input observed spectra provided a sufficient number of atomic lines are present in the stellar spectrum. In practice, this makes the {\sc LSDBinary} algorithm most suitable for stars of spectral type A and later. The regime of coolest stars whose spectra are dominated by molecular lines should be avoided as well for the reasons of the current lack of the algorithm validation in the respective parameter space;
    \item because resolving details of the line profile variations is an important aspect in the study of the in-eclipse spectra and Rossiter-McLaughlin effect, resolving power of the instrument should be sufficiently high for the purpose. In this study, we investigate the regimes of low ($R=5\,000$), medium ($R=25\,000$), and high ($R=65\,000$) resolution and find that the former of the three is largely insufficient irrespective of configuration of the binary system and amplitude of the Rossiter-McLaughlin effect. Depending on how large the latter is, a medium resolving power of the instrument of $R\approx25\,000$ might suffice, though we recommend to push for a higher resolution $R$ so that all details of the line profile variations can be resolved;
    \item the shapes of the obtained LSD profiles of both binary components are found by us to be largely insensitive to the variations in their atmospheric parameters \teff, \logg, and [M/H], provided these variations are within typical spectroscopic uncertainties. The largest effect is observed on the global scaling of the LSD profiles which leaves the RVs inferred from them unaffected. On the other hand, variations in \teff\ are found to have large effect on the components' radii ratio inferred with the {\sc LSDBinary} algorithm. This result is largely intuitive given that an incorrect assumption about \teff\ of the star leads to over- or under-estimation of its flux contribution to the total light of the system, the effect that is efficiently compensated in the {\sc LSDBinary} algorithm by variations of the stellar radius through the radii ratio parameter $R_2/R_1$. We note, however, that LSD profiles, and hence RVs inferred from them, are expected to be largely affected by non-uniform distributions of physical quantities (e.g. temperature, velocity, and/or chemical elements) over the stellar surface. The {\sc LSDBinary} code presented in this work includes a multiprofile LSD capability that, with just a mild development effort, could be exploited for calculation of multi-temperature and/or multi-chemical composition LSD profiles.\footnote{This particular capability is beyond the scope of this paper and is hence not elaborated on any further.};
    \item the orbital inclination angle $i$ and radii ratio $R_2/R_1$ are predictably the two parameters that have largest effect on the shape of the LSD profiles of the eclipsed binary component, and hence the shape and amplitude of the Rossiter-McLaughlin effect. Assuming orbital configuration of the binary system and one of the above parameters are known while the other one being optimised, the tandem of the {\sc RME} and {\sc LSDBinary} modules reconstructs successfully the pure orbital motion-driven RV curve of the eclipsed star. Should this not be the case, it is an indication of either incorrectly determined atmospheric parameters of either or both binary components, or of deficiencies in stellar atmosphere models that fail to predict correctly the components' flux ratio.
\end{itemize}

In practice, the latter finding can be exploited to either perform a consistency check between photometric and spectroscopic solutions of a binary system or to fine-tune the orbital inclination angle and/or components' radii ratio parameters from spectroscopic data. To demonstrate that, we return to the artificial in-eclipse data of an RZ\,Cas-like binary system whose configuration and atmospheric parameters are assumed to be known, except for the orbital inclination angle $i$ that needs a fine-tuning (see Table~\ref{Tab:RZCas_KIC11285625} for the list of the assumed parameters). We employ the tandem of the {\sc RME} and {\sc LSDBinary} codes to compute LSD profiles and RVs of both binary components from time-series of the simulated in-eclipse spectra, while keeping the radii ratio parameter $R_2/R_1$ free in the {\sc LSDBinary} module. The results are illustrated in the top and bottom panels in Figure~\ref{Fig:InclinationAngleOptimisation} for RV and $R_2/R_1$ as a function of orbital phase, respectively. Both curves are shown for three values of the inclination angle $i$: 76$^ {\circ}$ (open circles), 82$^ {\circ}$ (red stars), and 88$^ {\circ}$ (open squares). One can see that the best fit solution is achieved at the true value of the orbital inclination angle and corresponds to the RV curve resembling pure orbital motion of the star and the constant value of the radii ratio. This way, simultaneously minimising deviations of the in-eclipse RV curve from its pure orbital motion version and of the $R_2/R_1$ curve from constant value, allows one to fine-tune (or perform spectroscopic consistency check for) one of the parameters considered in this test.
  
We note that the {\sc LSDBinary} software package has been successfully applied to spectroscopic time-series of Algol-type systems with pulsating primary components, R\,CMa \citep{Lehmann2018} and RZ\,Cas \citep{Lehmann2020}. In both of these studies, the algorithm was used to compute the separated LSD profiles of both binary components and to measure precise orbital phase-resolved RVs from them. The separated LSD profiles of the primary component of the R\,CMa system were additionally used to study line profile variations caused by non-radial pulsations of the star \citep{Lehmann2018}.

\begin{acknowledgements}
The research leading to these results has (partially) received funding from the KU~Leuven Research Council (grant C16/18/005: PARADISE) and from the BELgian federal Science Policy Office (BELSPO) through PRODEX grant PLATO. Part of this work was supported by the German \emph{Deut\-sche For\-schungs\-ge\-mein\-schaft, DFG\/} project number Ts~17/2--1 and LE1102/3-1.
\end{acknowledgements}

\bibliographystyle{aa}
\bibliography{LSD}

\end{document}